\documentclass{LMCS}

\def\doi{8 (3:17) 2012}
\lmcsheading%
{\doi}
{1--37}
{}
{}
{Oct.~14, 2011}
{Sep.~17, 2012}
{}

\usepackage{enumerate}
\usepackage{amsmath}
\usepackage{amssymb}
\usepackage{stmaryrd}
\usepackage{graphicx}
\usepackage{listings}
\usepackage{microtype}
\usepackage{multirow}
\usepackage{multicol}
\usepackage{color}
\usepackage{algorithmic}
\usepackage{algorithm}
\usepackage{tikz}
\usepackage{hyperref}
\usetikzlibrary{trees}

\newcommand{\signed}[1]{\ensuremath{\langle\!\langle #1 \rangle\!\rangle}}
\newcommand{\unsigned}[1]{\ensuremath{\langle #1 \rangle}}

\def\mymin{\ell}
\def\mymax{u}

\theoremstyle{plain}

\def\vec#1{\mathchoice{\mbox{\boldmath$\displaystyle#1$}}
{\mbox{\boldmath$\textstyle#1$}}
{\mbox{\boldmath$\scriptstyle#1$}}
{\mbox{\boldmath$\scriptscriptstyle#1$}}}

\begin{document}

\title[Transfer Function Synthesis Without Quantifier Elimination]{Transfer Function Synthesis Without\\ Quantifier Elimination}

\author[J.~Brauer]{J{\"o}rg Brauer\rsuper a}	
\address{{\lsuper a}
Verified Systems International GmbH, Am Fallturm 1, 28359 Bremen, Germany
\and
Embedded Software Laboratory, RWTH Aachen University, Ahornstr. 55, 52074 Aachen, Germany}
\email{brauer@verified.de}

\author[A.~King]{Andy King\rsuper b}	
\address{{\lsuper b}Portcullis Computer Security Limited, Pinner, HA5 2EX, UK}
\email{a.m.king@kent.ac.uk} 







\begin{abstract}
  \noindent 
Traditionally, transfer functions have been designed manually for each operation in a program, instruction by instruction.
In such a setting, a transfer function describes the semantics of a single instruction, detailing how a given abstract input state is mapped to an abstract output state.
The net effect of a sequence of instructions, a basic block, can then be calculated by composing the transfer functions of the constituent instructions.
However, precision can be improved by applying a single transfer function that captures the semantics of the block as a whole.
Since blocks are program-dependent, this approach necessitates automation.
There has thus been growing interest in computing transfer functions automatically, most notably using techniques based on quantifier elimination.
Although conceptually elegant, quantifier elimination inevitably induces a computational bottleneck, which limits the applicability of these methods to small blocks.
This paper contributes a method for calculating transfer functions that finesses quantifier elimination altogether, and can thus be seen as a response to this problem.
The practicality of the method is demonstrated by generating transfer functions for input and output states that are described by linear template constraints, which include intervals and octagons.
\end{abstract}

\maketitle


\section{Introduction}
\label{section:introduction}

In model checking~\cite{BK08} the behaviour of a program
is formally specified with a model. Using the model,
all paths through the program are then
exhaustively checked against its requirements. 
The detailed nature of the requirements
entails that the program is simulated in a fine-grained way, sometimes
down to the level of individual bits. 
Because of the complexity of this reasoning
there has been much interest in 
abstracting away from the detailed nature of states. Then, the program
checker operates over classes of related states --- collections of
states that are equivalent in some sense --- rather than individual
states.

\subsection{Program analysis by abstract interpretation}

Abstract interpretation \cite{CC77} provides a systematic way to construct such
program checkers.  The key idea is
to simulate the execution of each concrete operation $g : C \to C$ in a program
with an abstract analogue $f : D \to D$ where $C$ and $D$ are
domains of concrete values and descriptions, respectively. Each abstract operation $f$ is designed to
faithfully model its concrete counterpart $g$ in the sense that
if $d \in D$ describes a concrete value $c \in C$,
sometimes written relationally as $d \propto c$~\cite{Mar93},
then the result of applying $g$ to $c$ is described by 
the action of applying $f$
to $d$, that is, $f(d) \propto g(c)$. Even for a fixed set of abstractions,
there are typically many ways of designing the abstract operations.
Ideally the abstract operations should compute abstractions that
are as descriptive, that is, as accurate as possible, though there
is usually interplay with accuracy and complexity, which is one
reason why the literature is so rich. Normally the abstract operations
are manually designed up front, prior to the analysis itself, but there
are distinct advantages in synthesising the abstract operations
from their concrete versions as part of the analysis itself, in a
fully automatic way, which is one reason why the topic is attracting increasing 
attention~\cite{BK10,BK11a,KS10,Mon09,Mon10b,regehr04hoist,RSY04} .

\subsection{The motivation for automatic abstraction}

One reason for automation stems from operations that arise in
sequences that are known as blocks.  Suppose that such a sequence
is formed of $n$ concrete operations $g_1, g_2, \ldots, g_n$, and each operation $g_i$ has its
own abstract counterpart $f_i$, henceforth referred to as its transfer
function~\cite{kam76global}. Suppose too that the input to the sequence $c \in C$ is described
by an input abstraction $d \in D$, that is, $d \propto c$.  Then the result of applying the $n$
concrete operations to the input (one after another) is described
by applying the composition of the $n$ transfer functions to the
abstract input, that is,
$f_n( \ldots f_2(f_1(d))) \propto g_n( \ldots g_2(g_1(c)))$.
However, a more accurate result can be obtained
by deriving a single
transfer function $f$ for the block $g_n \circ \ldots \circ g_2 \circ g_1$ as a whole,
designed so that
$f(d) \propto g_n( \ldots g_2(g_1(c)))$.
The value
of this approach has been demonstrated for linear congruences
\cite{Gra97} in the context of verifying
bit-twiddling code \cite{KS10}. 

To illustrate this interplay between
block-level abstraction and precision, consider
a block consisting of
three instructions
\mbox{x := y - x; y := y - x; x := x + y} that swaps the
values of the variables $x$ and $y$ without using a third variable~\cite[Chap.~2.19]{warren03hackers}.
To aid reasoning about the block as a whole, fresh variables are introduced,  static single assignment~\cite{CFR+91} style,
so as to separate different assignments to the same variable.  This gives
\mbox{$x''$ := $y - x$; $y'$ := $y - x''$; $x'$ := $x'' + y'$} where $x''$ is an intermediate and
$x$ and $x'$ (resp.~$y$ and $y'$) represent the values
of the variable $x$ (resp.~$y$) on entry and exit from the block. Since
\mbox{$x'' = y - x \wedge y' = y - x'' \wedge x' = x'' + y' \models y' = x \wedge x' = y$}
it follows that cumulatively the block
can be described by a pair of two variable equalities $x' = y \wedge y' = x$ which
can be interpreted as transfer function for the block. From this transfer function it follows that if $x = 1$ holds on entry
to the block then $y' = 1$ holds on exit. Note that equalities $x = 1$ and $y' = 1$ are
considered to be two-variable since they contain no more than two variables.
Now consider applying transfer functions for each of the
three assignments in turn. Again, suppose that $x = 1$ holds
prior to the
assignment $x'' := y - x$. Since the ternary constraint $x'' = y - 1$ cannot
be expressed within the two-variable equality domain then the best that can ever be inferred by
any transfer function operating over this domain is $x = 1$ for the post-state. Likewise
the best that can be inferred for a transfer function that simulates $y'$ := $y - x''$ is $x = 1$ for its post-state,
and similarly for $x'$ := $x'' + y'$.  Thus 
by composing transfer functions over two-variable equalities one cannot show that $y' = 1$ holds on exit from the block.
Therefore, the transfer function for a block
can be strictly more precise than the composition of the transfer functions
for the constituent instructions. Since blocks are program-dependent,
such an approach relies on automation rather than the manual provision
of transfer functions for each instruction.

Another compelling reason for automation is the complexity of the
concrete operations themselves; a problem that is heightened
by the finite nature of machine arithmetic.  For instance,
even a simple concrete operation,
such as increment by one, is complicated by the finite nature of
computer arithmetic: if increment is applied to the largest integer
that can be stored in a word, then the result is the smallest integer
that is representable. As the transfer function needs to faithfully
simulate concrete increment, the corner case inevitably manifests
itself (if not in the transfer function itself then elsewhere \cite{SK07b}).  

The problem of deriving transfer functions
for machine instructions, such as those of the x86, is particularly
acute \cite{balakrishnan07thesis} since these operations not only
update registers and memory locations, but also side effect status
flags~\cite{BHL+11,SMS11}, of which there are many.  When deriving
a transfer functions for 
a sequence of machine instructions
it is necessary to reason about how the status flags are used to
pass state from one instruction to another.
To illustrate the importance of status flags, consider double-length
addition, where the operands are pairs
of 32-bit words $(x_1, x_0)$ and $(y_1, y_0)$, the result is
denoted $(z_1, z_0)$, and the $1$ subscript denotes the most significant
half and $0$ the least significant. Then the following block 
$z_0 := x_0 + y_0; c := (z_0 < x_0); z_1 := x_1 + y_1 + c$
realises 64-bit addition, providing $<$ denotes an unsigned
comparison~\cite[Chap.~2.15]{warren03hackers}. Without considering the carry flag $c$, it is not clear
how one can reconstruct that
$(2^{32} \cdot z_1 + z_0) =  (2^{32} \cdot x_1 + x_0) + (2^{32} \cdot y_1 + y_0)$ modulo $2^{64}$
which is the high-level abstraction of the semantics of the block without resorting to 
a reduced cardinal power construction \cite[Theorem 10.2.0.1]{CC79}. 
In such a construction, a domain
that can express relations such as $z_0 < x_0$, henceforth called the base domain, is refined with respect to a
domain which traces the value of $c$, an adjunct that is sometimes called the exponent domain.  This refinement
enables $c$ to monitor whether $z_0 < x_0$ holds or not.
Although a base domain can always been refined in this way, and the transfer functions
enriched to support the extra expressiveness, an alternative approach is to
derive a transfer
function for a block of instructions which, in cases such as the above, better match against what can be 
expressed in the base domain.

As a final piece of motivation, it is worth noting that 
there are several ways of implementing double-length
addition, and numerous ways of realising other
commonly occurring operations \cite{warren03hackers}, and therefore
pattern matching can never yield a systematic nor
a reliable way of computing transfer functions for basic blocks.

\subsection{Specifying extreme values with universal quantifiers}

Monniaux \cite{Mon09,Mon10b} recently addressed the
vexing question of automatic abstraction by focussing on template
domains \cite{SSM04} which include, most notably,
intervals \cite{harrison77compiler} and octagons \cite{Min06}.  He
showed that if the concrete operations are specified as piecewise
linear functions, then it is possible to derive transfer functions
for blocks using quantifier elimination.  
To illustrate the role of quantification,
suppose a piecewise linear function models a block that updates three registers
whose values on entry and exit are represented by bit-vectors
$\vec{x}$, $\vec{y}$ and $\vec{z}$ 
and $\vec{x}'$, $\vec{y}'$ and $\vec{z}'$ respectively. 
To derive a transfer function for interval 
analysis, it is necessary to ascertain how the maximal value of $\vec{x}'$, 
denoted $\vec{x}'_{\mymax}$ say, relates to the minimal and maximal values 
of $\vec{x}$, $\vec{y}$ and $\vec{z}$, denoted 
$\vec{x}_{\mymin}$ and $\vec{x}_{\mymax}$,
$\vec{y}_{\mymin}$ and $\vec{y}_{\mymax}$
and
$\vec{z}_{\mymin}$ and $\vec{z}_{\mymax}$ respectively.
The value of $\vec{x}'_{\mymax}$ can be specified in logic~\cite{Mon09} 
by asserting that: 
\begin{iteMize}{$\bullet$}

\item
for all values of $\vec{x}$, $\vec{y}$ and $\vec{z}$
that fall within the intervals
$\vec{x} \in [\vec{x}_{\mymin},\vec{x}_{\mymax}]$,
$\vec{y} \in [\vec{y}_{\mymin},\vec{y}_{\mymax}]$
and
$\vec{z} \in [\vec{z}_{\mymin},\vec{z}_{\mymax}]$, the value of $\vec{x}'_{\mymax}$ 
is greater or equal to $\vec{x}'$

\item
for some combination of
values of $\vec{x}$, $\vec{y}$ and $\vec{z}$ such
that
$\vec{x} \in [\vec{x}_{\mymin},\vec{x}_{\mymax}]$,
$\vec{y} \in [\vec{y}_{\mymin},\vec{y}_{\mymax}]$
and
$\vec{z} \in [\vec{z}_{\mymin},\vec{z}_{\mymax}]$, the output $\vec{x}'$ takes the 
value of $\vec{x}'_{\mymax}$. 

\end{iteMize}
The ``for some'' can be expressed with 
existential quantification, and the ``for every''  with 
universal quantification. By applying quantifier elimination, a direct 
relationship between $\vec{x}_{\mymin}$, $\vec{x}_{\mymax}$, 
$\vec{y}_{\mymin}$, $\vec{y}_{\mymax}$,
$\vec{z}_{\mymin}$, $\vec{z}_{\mymax}$,
and 
$\vec{x}'_u$ can be found, yielding a mechanism for computing 
$\vec{x}'_u$ in terms of $\vec{x}_{\mymin}$, $\vec{x}_{\mymax}$, 
$\vec{y}_{\mymin}$, $\vec{y}_{\mymax}$,
$\vec{z}_{\mymin}$, $\vec{z}_{\mymax}$.
This construction is ingenious
but quantifier elimination is at least exponential for
rational and real piecewise linear systems
\cite{chandru03qualitative,weispfenning88complexity}, and is
doubly exponential when quantifiers alternate \cite{davenport88real}.
Hence, its application requires extreme care \cite{sturm11verification}.

As an alternative to operating over piecewise linear
systems~\cite{BK10}, one can instead express the semantics of a basic block
with a Boolean formula; an idea that is familiar in model checking
where it is colloquially referred to as bit-blasting~\cite{CKL04}.
First, bit-vector logic is used to
represent the semantics of a block as a single CNF formula $f_{\text{block}}$ (an excellent tutorial on flattening bit-vector logic into
propositional logic is given in~\cite[Chap. 6]{KS08b}).
Thus, each $n$-bit integer variable is represented as a separate
vector of $n$ propositional variables. Second, the above 
specification is applied to express the
maximal value (or conversely the minimal value) of an
output bit-vector in terms of the ranges on the input bit-vectors.
This gives a propositional formula $f_{\text{spec}}$ which is essentially
$f_{\text{block}}$ augmented with universal quantifiers and existential quantifiers.
Third, the quantifiers are
removed from $f_{\text{spec}}$ to obtain $f_{\text{simp}}$ -- a 
simplification of $f_{\text{spec}}$.  
Of course, $f_{\text{simp}}$ is just a Boolean formula and does not prescribe how
to compute a transfer function.  However, a transfer function
can be extracted from $f_{\text{simp}}$ by abstracting $f_{\text{simp}}$ with
linear affine equations \cite{Kar76} which directly relate the output ranges
to the input ranges. This fourth step (which is
analogous to that proposed for abstracting formulae with
congruences \cite{KS10})
is the final step in the construction.

This proposal for computing transfer functions \cite{BK10} may seem attractive
since computing $\forall \vec{y} : \varphi$,
where $\varphi$ is a system 
of propositional constraints and $\vec{y}$ is a vector of variables, is 
straightforward when the formula $\varphi$ is in CNF.
When $\varphi$ is an arbitrary propositional system, a CNF formula $\psi$ that is 
equisatisfiable, denoted $\equiv$, to $\varphi$ can be found~\cite{PG86} 
by introducing fresh variables $\vec{z}$ to
give $\varphi \equiv \exists \vec{z} : \psi$. However, then the
transfer function synthesis problem amounts to
solving $\forall \vec{y} : \exists \vec{z} : \psi$ where $\psi$ is in CNF. 
To eliminate the existentially quantified variables $\vec{z}$,
resolution~\cite[Chap.~9.2.3]{KS08b} can be applied, but the quadratic nature of each resolution step compromises tractability as the size of $\vec{z}$ increases. The size of $\vec{z}$ is proportional to the number of logical connectives in $\varphi$ which, in turn, depends on the size of the bit-vectors and the complexity of the block under consideration. It is therefore no surprise that this approach has only been demonstrated for blocks of microcontroller code where the word-size is just 8 bits~\cite{BK10}. Although no polynomial-time algorithms are known for existential quantifier elimination of CNF, new algorithms are emerging
\cite{BKK11a} which will no doubt permit transfer functions to be derived for larger blocks.
Nevertheless, it would be preferable if quantifier elimination was avoided altogether.

\subsection{Avoiding quantifier elimination}

This paper develops the work reported in \cite{BK11a} to contribute a method
for deriving transfer functions which replaces
quantifier elimination
with successive calls to a SAT solver, where the number
of calls grows linearly with the word-size rather than the size of the
formula that encodes the semantics of the block.

To illustrate, consider an octagon~\cite{Min06}
which consists of a system of inequalities
of the form $\pm x \pm y \leq d$. For each of these inequalities,
our approach derives the least $d \in \mathbb{Z}$ (which is uniquely 
determined) such that the inequality holds for all feasible values of $x$ and $y$
as defined by some propositional formula.
As an example, consider 
the inequality $x + y \leq d$. The constant $d$ is defined as 
$d = \min \{ c \in \mathbb{Z} \mid \forall \vec{x} : \forall \vec{y} : f(\vec{x},\vec{y}) \wedge \vec{x} + \vec{y} \leq c \}$ 
where $f(\vec{x},\vec{y})$ is a propositional formula constraining the 
bit-vectors $\vec{x}$ and $\vec{y}$. Furthermore, given a machine 
with word-length $w$, the maximal value in an unsigned representation of $x$ and $y$
is $2^w-1$, and thus we can derive an initial constraint 
$0 \leq d \wedge d \leq 2 \cdot (2^w-1)$ for $d$, which can be 
expressed disjunctively as $\mu_{\mymin} \vee \mu_{\mymax}$ where:
\begin{iteMize}{$\bullet$}

\item
$\mu_{\mymin} = 0 \leq d \wedge d \leq 2^w-1$

\item
$\mu_{\mymax} = 2^w \leq d \wedge d \leq 2 \cdot( 2^w-1)$

\end{iteMize}
To determine which disjunct characterises $d$, it is sufficient to test
the propositional formula $\exists \vec{x} : \exists \vec{y} : f(\vec{x},\vec{y}) \wedge \vec{x}+\vec{y} \geq 2^w$ for \emph{satisfiability}. If satisfiable, then $\mu_{\mymax}$
is entailed by the inequality $x + y \leq d$, and $\mu_{\mymin}$ otherwise. We proceed by decomposing
the new characterisation into
a disjunction --- as in dichotomic or binary search --- and repeating this step $w$ times to give $d$ exactly.  
Likewise, constants $d$ can be found for
all inequalities of the form $\pm x \pm y \leq d$, which provides a mechanism for
computing an octagonal abstraction that describes a given propositional formula.
The force of this abstraction technique is that it provides a way of
deriving octagonal guards
which must hold for a block to be executed in a particular mode.  
For example, a block might have three modes of operation, depending
on whether an operation underflows, overflows, or does neither.
Which mode is applicable then depends on the
values of variables on entry to the block, which motivates using
guards to separate and describe the different modes of operation.
Knowing that a particular mode is applicable permits 
a specialised transfer function to be applied for inputs that conform to that mode.
It is important to note that separating modes is a crucial step in the
process of applying abstract domains that operate on unbounded integers,
such as affine equalities, to describe finite bit-vector semantics.
As an example, consider incrementing a variable $x$
by 1. If $x$ and its representative $x'$ on output are unbounded
integers, the affine relation is merely $x' = x+1$. Now suppose that  $x$
and $x'$ are 32-bit variables. Then, if $x < 2^{32}-1$ it follows $x' = x+1$, and
$x' = 0$ otherwise. Even though
each of the two cases can be described in the affine domain, the
join of these two affine relations conveys no useful information at all.
Separating modes ultimately leads to a transfer function being formulated as a system
of guarded updates, where the updates stipulate how the entry values are mapped
to exit updates, and the guards indicate which mode holds and therefore which type
of update is applicable.

This leaves the problem of how to compute the updates themselves; the
input-output transformers that
constitute the heart of the transfer function. We show that updates
can be also computed without resorting to quantifier elimination.   We demonstrate
this construction not only for intervals, but for transfer functions
over octagons. The method is based on computing an
affine abstraction of a Boolean formula that is derived to describe the mode.
For intervals, the update details
how the bounds of an input interval are mapped to new bounds of
an output interval. For octagons, the update maps the
constants on the input octagonal inequalities to new constants on
the output inequalities.  

\subsection{Contributions}

Overall, the approach to  computing transfer functions that is presented in this paper
confers the following advantages:
\begin{iteMize}{$\bullet$}

\item
it is amenable to instructions whose semantics is presented
as propositional formulae or Satisfiability Modulo Theory (SMT) \cite{smtsurvey} formulae.  The force of this is that
such encodings are
readily available for instructions, due to the
rise in popularity of SAT-based model checking;

\item
it avoids the computational problems associated with
eliminating variables from piecewise linear systems and
propositional formulae, particularly with regard to
alternating quantifiers;

\item
it proposes the use of transfer functions that are 
action systems of guarded updates.  These transfer
functions are attractive both in terms of their
expressiveness and the ease with which they can be
evaluated (only one expression need be evaluated for each
inequality that describes the state on exit from the block);

\item
it shows how the modes of a block can be found and how, for a given mode, 
the guards can be computed using repeated SAT solving.  It is also shown how the
updates for that mode
can be deriving by interleaving SAT solving with affine abstraction;

\item
it shows how update operations, which in the case of interval
analysis, compute bounds on the output intervals from
bounds on the
input intervals, need not be linear functions. Non-linear
update operations can also be supported for transfer functions
over octagons.  In this context, the update  
operation computes the constants
on the output octagonal inequalities from the constants on the input inequalities (the coefficients are
fixed in both the input and output octagons hence computing a transfer function amounts to adjusting constants);

\item
it explains how to handle operations that underflow, overflow, or do neither and even
combinations of such behaviours, providing a way to seamlessly integrate 
template inequalities with finite precision arithmetic.


\end{iteMize}



\section{Outline of the approach}

Overall the paper proposes a systematic technique for inferring transfer
functions that are defined as systems of guarded updates.  This section illustrates
the syntactic form of transfer functions, so as to
provide an outline of the approach and a roadmap for the whole paper.
The roadmap explains which sections of the paper
are concerned with deriving which components of the transfer function.

\subsection{Modes}
Transfer functions are inferred for blocks, such as the assembly code
listing in Fig.~\ref{figure:example_program}.
(The approach is illustrated for blocks of 32-bit AVR UC3
assembly code~\cite{UC3}, though the techniques are completely generic.)
Each instruction is modelled by at least one, and at most four, 
Boolean functions according to whether it overflows or underflows, or
is exact, that is, whether the instruction neither overflows nor underflows.  This
division into three cases reflects the ways the two's complement overflow (V) flag is set or clearer \cite{UC3}.  In
exceptional cases this flag is used in tandem with the negative (N) flag \cite{UC3} and thus it is natural
to refine these three cases according to whether the negative flag is also set or clearer. However,
if the instruction overflows 
then the result is necessarily negative whereas if it underflows then the result is non-negative,
hence only the exact case needs to be further partitioned. This
gives four cases in all, overflow, underflow, exact and negative, exact and non-negative,
each of which can be precisely expressed with
a Boolean function that describes a so-called \emph{mode}.
The different instructions that make up the block
may operate in different modes, though the mode of one instruction may preclude a mode
of another being applicable.
A mode combination is
then chosen for each instruction, and a single Boolean formula is constructed for the block by composing a
formula for each instruction in the prescribed mode. If the composed formula is unsatisfiable, then the
mode combination is inconsistent.  Otherwise, the mode combination is feasible and the formula describes one
type of wrapping (or non-wrapping) behaviour that can be realised within the block. 

\subsection{Transfer functions}
The composed formula is
then used to distill a guard paired with an update; one pair is computed for each feasible
mode combination.  For example, the block listed in 
Fig.~\ref{figure:example_program} has nine feasible mode combinations in all, yielding
nine guard and update pairs of the form:

\[
\begin{array}{@{}r@{\;}c@{\;}l@{}}
\left. \begin{array}{@{}rr}
2^{31} \leq \signed{\vec{r0}} + \signed{\vec{r1}} \leq 2^{31} & \wedge \\
1 \leq \signed{\vec{r0}} \leq 2^{31}-1 & \wedge \\
1 \leq \signed{\vec{r1}} \leq 2^{31}-1 \\
\end{array} \right\} 
&
\rightarrow
&
\left\{
\begin{array}{l@{\;}l@{\;}ll@{}}
(\signed{\vec{r0}_{\mymin}'} & = & -2^{31}) & \wedge \\ 
 (\signed{\vec{r0}_{\mymax}'} & = & -2^{31})
\end{array}\right.
\\[3ex]
\left. \begin{array}{@{}rr}
2^{31} + 1 \leq \signed{\vec{r0}} + \signed{\vec{r1}} \leq 2^{32} - 2 & \wedge \\
0 \leq \signed{\vec{r0}} \leq 2^{31}-1 & \wedge \\
0 \leq \signed{\vec{r1}} \leq 2^{31}-1 
\end{array} \right\}
&
\rightarrow
&
\left\{
\begin{array}{l@{\;}l@{\;}ll@{}}
(\signed{\vec{r0}_{\mymin}'} & = & 2^{32} - \signed{\vec{r0}_{\mymax}} - \signed{\vec{r1}_{\mymax}}) & \wedge \\ 
 (\signed{\vec{r0}_{\mymax}'} & = & 2^{32} - \signed{\vec{r0}_{\mymin}} - \signed{\vec{r1}_{\mymin}})
\end{array}\right.
\\[3ex]
& \vdots & 
\\
\left. \begin{array}{@{}rr} 
-2^{31} +1 \leq \signed{\vec{r0}} + \signed{\vec{r1}} \leq -1 & \wedge \\
0 \leq \signed{\vec{r1}} \leq 2^{31} - 1
\end{array} \right\} 
&
\rightarrow
&
\left\{
\begin{array}{l@{\;}l@{\;}ll@{}}
(\signed{\vec{r0}_{\mymin}'} & = & -\signed{\vec{r0}_{\mymax}} - \signed{\vec{r1}_{\mymax}}) & \wedge \\ 
 (\signed{\vec{r0}_{\mymax}'} & = & -\signed{\vec{r0}_{\mymin}} - \signed{\vec{r1}_{\mymin}})
\end{array}\right.\\[2ex]
\end{array} 
\]

\noindent
Each guard is a conjunction of linear template constraints over the inputs
of the block, in this case $\signed{\vec{r0}}$ and $\signed{\vec{r1}}$, which denote
the (signed) values of the registers \texttt{R0} and \texttt{R1} on entry to the block.  The guards
express properties of \texttt{R0} and \texttt{R1} which must hold for the instructions to
operate in the modes that make up the feasible mode combination.

The update operations that augment the guards detail how the values of the registers
are mutated for a given mode combination.  For example, if the first guard is applicable, then the update
asserts that the output value
of \texttt{R0} takes a value in the range  $[-2^{31}, -2^{31}]$ (which actually prescribes a single value); the lower and upper bounds
of \texttt{R0} on exit are denoted $\signed{\vec{r0}_{\mymin}'}$
and $\signed{\vec{r0}_{\mymax}'}$ in the update.  The second update illustrates how 
$\signed{\vec{r0}_{\mymin}'}$
and $\signed{\vec{r0}_{\mymax}'}$ can depend on the values of \texttt{R0} and
\texttt{R1} on entry to the block, where the input lower and upper bounds for  \texttt{R0} are denoted 
$\signed{\vec{r0}_{\mymin}} $ and $\signed{\vec{r0}_{\mymax}} $, and likewise for \texttt{R1}.

\subsection{Automatic derivation and roadmap}
The guard is constructed one inequality at a time, by applying a form of dichotomic (or binary)
search. This step amounts to a series of calls to a SAT solver, as is explained in
Sect.~\ref{section:guards}. Updates can be computed
by inferring an affine relationship between the bound on an output symbolic constraint and the input
symbolic bounds. Such a relationship can again be derived by repeated SAT solving, as
detailed in Sect.~\ref{section:updates}. 
Replicating this construction for each of the symbolic output constants
gives the update operation for a feasible mode combination. (Sect.~\ref{section:guards} and
Sect.~\ref{section:updates} return
to the example introduced given above, detailing the steps in the derivation of this transfer function.)
Yet, situations can arise for which the updates cannot be expressed using
affine relationships, motiving the study, in Sect.~\ref{section:template_octahedra},
of complementary classes of update which can
be formed from linear template inequality constraints~\cite{CC04} and
non-linear template equality constraints~\cite{Col04}.
Updates that relate symbolic output constants to symbolic input constraints using
equalities are complementary
to those based on inequalities: both are useful when transfer functions are
evaluated. Sect.~\ref{section:evaluating} focuses on this topic and explains how guards and updates are applied during
fixed point computation.
Evidence is presented in Sect.~\ref{section:experiments} which demonstrates that
the techniques presented in the paper are capable of
synthesising transfer functions for blocks, where previous
approaches based on quantifier elimination were prohibitively expensive.
Finally, Sect.~\ref{section:related_work} surveys the related work and Sect.~\ref{section:discussion}
concludes.



\section{Deriving Guards}
\label{section:guards}

We express the concrete semantics of a block with Boolean formulae
so as to ultimately infer a set of guards that distinguish that wrapping
behaviour of a block. The construction given in~\cite{BK10} formulates
this problem using quantification, so that quantifier elimination can
be applied to solve it.  However, whereas universal quantifier elimination
is attractive computationally, this is not so for the  elimination of existentially
quantified variables.  We overcome this problem by reformulating the construction given
in~\cite{BK10}, and replace quantifier elimination by a series of calls to a SAT solver.
This section illustrates the power of this transposition by deriving guards for some
illustrative blocks of microcontroller instructions.

%
%

\subsection{Deriving interval guards by range refinement}
\label{section:worked_examples:deriving_interval_guards}

Consider deriving a transfer function for the operation \texttt{INC R0}, 
which increments the value of \texttt{R0} by one and stores the result 
in \texttt{R0}. For this example, we assume that the operands are
unsigned. We represent the value of \texttt{R0} by a bit-vector
$\vec{r0}$ and let $\unsigned{\vec{r0}} = \sum_{i=0}^{31} 2^i \cdot \vec{r0}[i]$ 
where $\vec{r0}[i]$ denotes the $i^{\text{th}}$ element of $\vec{r0}$. 
Note that in the sequel the following notational distinction is maintained:
\texttt{R0} for a register,
$\vec{r0}$ for a bit-vector representing \texttt{R0} and
$\unsigned{\vec{r0}}$ and
$\signed{\vec{r0}}$ for, respectively, the unsigned 
and signed interpretation of the bit-vector $\vec{r0}$.
The instruction itself can operate in one of two modes: (1) it overflows 
(iff $\unsigned{\vec{r0}} = 2^{32}-1$) or (2) it is exact (otherwise). 
Note that in the sequel the term exact is used 
to refer to a mode that is neither underflowing nor overflowing.
The
semantics of these two modes can be expressed as two formulae:
\[
\begin{array}{lllllllllll}
(1) & \hspace{0.3em} & \varphi_O(\vec{X}) & = & \varphi(\vec{X}) \wedge (\bigwedge_{i=0}^{31} \vec{r0}[i]) \\
(2) & \hspace{0.3em} & \varphi_E(\vec{X}) & = & \varphi(\vec{X}) \wedge ( \bigvee_{i=0}^{31} \neg \vec{r0}[i])
\end{array}
\]
where $\varphi(\vec{X})$ encodes the increment over bit-vectors
$\vec{X} = \{\vec{r0},\vec{r0}' \}$ as follows:
\[
\begin{array}{lll}
\varphi(\vec{X}) & = & \bigwedge_{i=0}^{31} \left(\vec{r0}'[i] \leftrightarrow \vec{r0}[i] \oplus \bigwedge_{j=0}^{i-1} \vec{r0}[j]\right)
\end{array}
\]
Both formulae can be converted into equisatisfiable formulae in CNF by
introducing fresh variables $\vec{z}$~\cite{PG86,Tse68}.
We therefore denote the resulting formulae by $\varphi_E(\vec{X},\vec{z})$ and 
$\varphi_O(\vec{X},\vec{z})$. Following our initial approach~\cite{BK10}, the 
transfer function for a multi-modal block (where the internal instructions can 
wrap) is described as a system of guarded updates. In the one-dimensional 
case, octagonal guards coincide with intervals. Each guard constitutes an 
upper-approximation of those inputs that are compatible with the specific 
mode. In case of the increment, we derive guards $g_O$ and $g_E$
defined as:
\[
\begin{array}{lllllllll}
(1) & \hspace{0.2em} & g_O & = & 2^{32}-1 & \leq & \unsigned{\vec{r0}} & \leq & 2^{32}-1 
\\[0.2ex]
(2) & \hspace{0.2em} & g_E & = & 0 & \leq & \unsigned{\vec{r0}} & \leq & 2^{32}-2
\end{array}
\]
These guards partition the inputs into two disjoint spaces: (1) a single point for the overflow case and
(2) exact operation. To obtain these guards, we provide an algorithm which solves
a series of SAT instances, rather than
following a monolithic all-in-one approach based on quantifier
elimination~\cite{BK10}. To illustrate our strategy, consider the computation
of a least upper bound $d$ for $\unsigned{\vec{r0}}$ for the formula
$\varphi_E(\vec{X},\vec{z})$. Clearly, we have $0 \leq d \leq 2^{32}-1$. We start by putting:
\[
\begin{array}{lll}
\psi^1_E(\vec{X},\vec{z}) & = & \varphi_E(\vec{X},\vec{z}) \wedge \unsigned{\vec{r0}} \geq 2^{31}
\end{array}
\]
Recall that we use a binary encoding of integers in the Boolean formulae. Further,
as $2^{31}$  is a power of two, we can finesse the need for a complicated Boolean encoding of the
predicate $\unsigned{\vec{r0}} \geq 2^{31}$ by using the equivalent formula: 
\[
\begin{array}{lll}
\psi^{\mathsf{simp},1}_E(\vec{X},\vec{z}) & = & \varphi_E(\vec{X},\vec{z}) \wedge \vec{r0}[31]
\end{array}
\]
which is simpler both to formulate and to solve. Then, the satisfiability of 
$\psi^{\mathsf{simp},1}_E(\vec{X},\vec{z})$ shows that $\vec{r0}$ takes
a value in the range $2^{31} \leq \unsigned{\vec{r0}} \leq 2^{32}-1$. Consequently, 
$d$ occurs in the same range. We can thus further refine this range by testing:
\[
\begin{array}{lll}
\psi^2_E(\vec{X},\vec{z}) & = & \varphi_E(\vec{z}) \wedge \unsigned{\vec{r0}} \geq (2^{31} + 2^{30})
\end{array}
\]
for satisfiability, or equivalently: 
\[
\begin{array}{lll}
\varphi^{\mathsf{simp},2}_E(\vec{X},\vec{z}) & = & \varphi_E(\vec{z}) \wedge \vec{r0}[31] \wedge \vec{r0}[30]
\end{array}
\]
As $\psi^{\mathsf{simp},2}_E(\vec{X},\vec{z})$ is satisfiable, we infer that $d$
satisfies $2^{30} + 2^{31} \leq d \leq 2^{32}-1$. The method continues to refine 
the constraint on $d$ into two equally sized halves. Only in the last iteration is the 
satisfiability check found to fail, from which we conclude that 
$d = \sum_{i=1}^{31} 2^i = 2^{32}-2$. 
Overall, this deduction requires 32 
SAT instances, but the similarity of the instances suggests that the overhead can 
be mitigated somewhat by incremental SAT.

\subsection{Deriving octagonal guards by range refinement}
\label{section:worked_examples:deriving_octagon_guards}

\begin{figure}[t]\label{figure:example_program}
\[
\begin{array}{r@{\;}l@{\qquad}r@{\;}l@{\qquad}r@{\;}l@{\qquad}r@{\;}l}
1: & \texttt{ADD R0 R1}; &
2: & \texttt{MOV R2 R0}; &
3: & \texttt{EOR R2 R1}; &
4: & \texttt{LSL R2}; \\
5: & \texttt{SBC R2 R2}; &
6: & \texttt{ADD R0 R2}; &
7: & \texttt{EOR R0 R2}; & &
\end{array}
\]
\caption{Assembly listing corresponding to the assignment \texttt{R0' := isign(R0+R1,R1)}}
\end{figure}

In a second example, we show how to extend the refinement
technique from intervals to octagons. To illustrate the method, 
consider the program fragment in Fig.~\ref{figure:example_program}.
This program corresponds to an assignment \mbox{\texttt{R0' := isign(R0+R1,R1)}} 
for signed values. The function \texttt{isign} assigns \texttt{abs(R0+R1)} to 
\texttt{R0} if \texttt{R1} is positive, and \texttt{-abs(R0+R1)} otherwise. 
\texttt{R2} is used as a temporary register. The sum of \texttt{R0} and 
\texttt{R1} is computed by instruction $(1)$, and instructions $(2)$ -- $(7)$
implement \texttt{isign}. The semantics of even this simple block is not obvious
due to the bounded nature of machine arithmetic. For instance, if
\texttt{abs} is applied to the smallest representable integer $-2^{31}$ 
then the result is $2^{31}$ subject to overflow, which gives $-2^{31}$. 
To derive octagons that describe such corner cases, we have to consider 
all combinations of over- and underflow modes of the instructions. In the 
above program, the instructions \texttt{ADD} (sum) and \texttt{LSL} (left-shift) 
can wrap in different ways, and thus are multi-modal. Neither \texttt{EOR} 
nor \texttt{MOV} can wrap; they are both  uni-modal. Note that in general, 
the instruction \texttt{SBC}  (subtract-with-carry) is multi-modal. However, in the
 case of two equal operands, the instruction can only result in $0$ or $-1$, depending 
on the carry-flag.  We thus ignore the wrapping of
\texttt{SBC R2 R2} and consider it to be uni-modal for simplicity of presentation.  Note that only
overflows occurred in the previous example since the  single operand was unsigned.

\subsubsection{Finding the feasible mode combinations}


In what follows, let $\mu(\vec{X})$ defined as
\[
\begin{array}{cl}
\mu(\vec{X}) = & \left(\bigwedge_{i=0}^{31} \vec{r0}'[i] \leftrightarrow \vec{r0}[i] \oplus
  \vec{r1}[i] \oplus \vec{c}[i] \right) \wedge \\
& \neg \vec{c}[0] \wedge
\left( \bigwedge_{i=0}^{30} \vec{c}[i+1] \leftrightarrow (\vec{r0}[i] \wedge \vec{r1}[i]) \vee (\vec{r0}[i] \wedge \vec{c}[i]) \vee (\vec{r1}[i] \wedge \vec{c}[i]) \right)
\end{array}
\]
denote the Boolean encoding of the instruction \mbox{\texttt{ADD R0 R1}} over bit-vectors
$\vec{X} = \{ \vec{r0},\vec{r1}, \dots \}$ obtained through static single assignment conversion.
Here, $\vec{c}$ is a bit-vector of intermediate carry bits. The semantics of \texttt{ADD R0 R1}
is to compute the sum of \texttt{R0} and \texttt{R1} and store the result in \texttt{R0}. Since we
are now working with signed objects, let
\[
\begin{array}{lll}
\signed{\vec{x}} & = & (\sum_{i=0}^{w-2} 2^i \cdot \vec{x}[i]) - 2^{w-1} \cdot \vec{x}[w-1]
\end{array}
\]
denote the value of a bit-vector $\vec{x}$ of length $w$, where $\vec{x}[w-1]$ is interpreted as the sign-bit. Then,
\texttt{ADD R0 R1} has four modes of operation: overflow, underflow, exact and non-negative, exact and negative. 
Underflow occurs, for example, if the arithmetic sum of $\signed{\vec{r0}}$ and $\signed{\vec{r1}}$
is less than $-2^{31}$. The constraints for these modes, which are obtained directly from the
instruction set specification~\cite[p.~127]{UC3}, can be expressed as four Boolean formulae:
\[
\begin{array}{lll}
\mu_{O}(\vec{X}) & = & \neg \vec{r0}[31] \wedge \neg \vec{r1}[31] \wedge \vec{r0}'[31] \\
\mu_{U}(\vec{X}) & = & \vec{r0}[31] \wedge \vec{r1}[31] \wedge \neg \vec{r0}'[31] \\
\mu_{P}(\vec{X}) & = & (\vec{r0}[31] \vee \vec{r1}[31] \vee \neg \vec{r0}'[31]) \wedge (\neg \vec{r0}[31] \vee \neg \vec{r1}[31] \vee \vec{r0}'[31]) \wedge \neg \vec{r0'}[31] \\
& = & (\neg \vec{r0}[31] \vee \neg \vec{r1}[31] \vee \vec{r0}'[31]) \wedge \neg \vec{r0'}[31] \\
& = & (\neg \vec{r0}[31] \vee \neg \vec{r1}[31]) \wedge \neg \vec{r0'}[31] \\
\mu_{N}(\vec{X}) & = & (\vec{r0}[31] \vee \vec{r1}[31] \vee \neg \vec{r0}'[31]) \wedge (\neg \vec{r0}[31] \vee \neg \vec{r1}[31] \vee \vec{r0}'[31]) \wedge \vec{r0'}[31] \\
& = & (\vec{r0}[31] \vee \vec{r1}[31] \vee \neg \vec{r0}'[31]) \wedge \vec{r0'}[31] \\
& = & (\vec{r0}[31] \vee \vec{r1}[31] ) \wedge \vec{r0'}[31] \\
\end{array}
\]
For example, the formula $\mu(\vec{X}) \wedge \mu_O(\vec{X})$ describes the input-output relationships
for \mbox{\texttt{ADD R0 R1}} in overflow mode. The instruction \texttt{LSL R2} shifts register \texttt{R2} to the left by one bit-position;
the most-significant bit of \texttt{R2} is moved into the carry-flag. If the carry-flag is set,
an overflow occurs; there is no underflow for \texttt{LSL}. Let $\nu(\vec{X}) \wedge \nu_O(\vec{X})$ and
$\nu(\vec{X}) \wedge \nu_E(\vec{X})$ thus express the overflow
and exact modes of \texttt{LSL R2}. In an analogous way to the first \texttt{ADD} instruction,
let $\eta(\vec{X}) \wedge \eta_O(\vec{X})$, $\eta(\vec{X}) \wedge \eta_U(\vec{X})$,
$\eta(\vec{X}) \wedge \eta_P(\vec{X})$
and $\eta(\vec{X}) \wedge \eta_N(\vec{X})$ express the semantics of
the instruction \mbox{\texttt{ADD R0 R2}}. Using these encodings that satisfy a single mode, we
can compose a Boolean formula for a fixed mode combination that expresses the possibility 
of one mode of one operation being consistent with another mode of another operation;
the unsatisfiability of this formula indicates that the chosen modes are inconsistent. For
example, the combination of $\mu_U(\vec{X})$, $\nu_E(\vec{X})$ and $\eta_P(\vec{X})$ is
infeasible. The above block thus constitutes $4 \cdot 2 \cdot 4 = 32$ combinations of modes, but
only $9$ of which are satisfiable, which is depicted in Tab.~\ref{table:feasible_modes}. It is thus
necessary to derive guards only for the feasible combinations.


\begin{table}
\caption{Feasible and infeasible modes for the program in Fig.~\ref{figure:example_program}}
\label{table:feasible_modes}
\begin{tabular}{@{}c@{}||@{}c@{}}
\begin{tabular}{@{}c|c|c|c|@{}}
\hline
\texttt{ADD R0 R1} & \texttt{LSL R2} & \texttt{ADD R0 R2} & feasible? \\
\hline
\hline
O & E & O & no \\
O & E & U & no \\
O & E & P & no \\
O & E & N &Êno \\
O & O & O & no \\
O & O & U & \textbf{yes} \\
O & O & P & no \\
O & O & N & \textbf{yes} \\
U & E & O & no \\
U & E & U & no \\
U & E & P & no \\
U & E & N &Êno \\
U & O & O & no \\
U & O & U & no \\
U & O & P & \textbf{yes} \\
U & O & N & \textbf{yes} \\
\hline
\end{tabular}
&
\begin{tabular}{@{}|c|c|c|c@{}}
\hline
\texttt{ADD R0 R1} & \texttt{LSL R2} & \texttt{ADD R0 R2} & feasible?~ \\
\hline
\hline
P & E & O & no \\
P & E & U & no \\
P & E & P & \textbf{yes} \\
P & E & N &Êno \\
P & O & O & no \\
P & O & U & no \\
P & O & P & \textbf{yes} \\
P & O & N & no \\
N & E & O & no \\
N & E & U & no \\
N & E & P & no \\
N & E & N &Ê\textbf{yes} \\
N & O & O & no \\
N & O & U & \textbf{yes} \\
N & O & P & no \\
N & O & N & \textbf{yes} \\
\hline
\end{tabular}
\end{tabular}
\end{table}

\subsubsection{Incremental elimination of mode combinations}

The number of mode combinations in a single basic block is, in the worst case, exponential
in the number of instructions in the block. The number of calls to a SAT solver
required to determine feasibility is thus exponential too. Further, incremental SAT solving \cite{WKS01}, which greatly affects the efficiency
of modern solvers, cannot be exploited when the feasibility of the mode combinations are checked one-by-one. 
We therefore present a strategy for incrementally
checking the feasibility of mode combinations.
To illustrate, let $\varphi(\vec{X})$ encode the instructions of the entire block and consider the
case where \texttt{ADD R0 R1} underflows and \texttt{LSL R2} is exact. The formula
\[
\varphi(\vec{X})Ê\wedge \mu_U(\vec{X}) \wedge \nu_E(\vec{X})
\]
describes this compound mode, independent of the second \texttt{ADD}.
Since this formula is unsatisfiable
is follows that the mode combinations 
\[
\begin{array}{l}
\varphi(\vec{X})Ê\wedge \mu_U(\vec{X}) \wedge \nu_E(\vec{X}) \wedge \eta_O(\vec{X}), \\
\varphi(\vec{X})Ê\wedge \mu_U(\vec{X}) \wedge \nu_E(\vec{X}) \wedge \eta_U(\vec{X}), \\
\varphi(\vec{X})Ê\wedge \mu_U(\vec{X}) \wedge \nu_E(\vec{X}) \wedge \eta_P(\vec{X}), \\
\varphi(\vec{X})Ê\wedge \mu_U(\vec{X}) \wedge \nu_E(\vec{X}) \wedge \eta_N(\vec{X}) \\
\end{array}
\]
are also  infeasible.
This suggests extending the formula $\varphi(\vec{X})$ with mode constraints, such as $\mu_U(\vec{X})$,
in a tree-like fashion, instruction by instruction. A sub-tree, which represents a different modes of one
instruction, is then created and followed iff the formula is satisfiable. This strategy is illustrated
in Fig.~\ref{figure:sat_tree}. Observe that the technique may increase the overall number of SAT instances to
be solved: 36 instead of 32 for the running example. In the worst case, if all leaves are reachable,
the strategy requires an exponential number of SAT calls. However, the tree-like strategy integrates smoothly
with incremental SAT solving~\cite{WKS01} since the additional mode constraints can be passed as assumptions, thereby
permitting an incremental SAT solver to reuse learnt clauses. Which technique outperforms the other strongly depends
on the distribution of feasible modes, there is thus no clear winner.

\tikzstyle{level 1}=[level distance=4cm, sibling distance=5cm]
\tikzstyle{level 2}=[level distance=4cm, sibling distance=2.5cm]
\tikzstyle{level 3}=[level distance=4cm, sibling distance=0.57cm]

\begin{figure}[!htb]
\small
\begin{tikzpicture}[grow=right,sloped]
\node{$\varphi$}
        child {
        node{SAT}        
        child {
                node{SAT}
        		child {
                		node{\textbf{SAT}}
                		edge from parent
                		node[above] {$\eta_N$}
            	}
        		child {
                		node{UNSAT}
		    	edge from parent
                		node[above] {$\eta_P$}
            	}
            	child {
                		node{UNSAT}
                		edge from parent
                		node[above] {$\eta_U$}
            	}
	         child {
                		node{UNSAT}
                		edge from parent
                		node[above] {$\eta_O$}
            	}
                edge from parent
                node[above] {$\nu_E$}
            }
        child {
                node{SAT}
        		child {
                		node{\textbf{SAT}}
                		edge from parent
                		node[above] {$\eta_N$}
            	}
        		child {
                		node{UNSAT}
		    	edge from parent
                		node[above] {$\eta_P$}
            	}
            	child {
                		node{\textbf{SAT}}
                		edge from parent
                		node[above] {$\eta_U$}
            	}
	         child {
                		node{UNSAT}
                		edge from parent
                		node[above] {$\eta_O$}
            	}
                edge from parent
                node[above] {$\nu_O$}
            }
        edge from parent         
            node[above] {$\mu_N$}
    }
        child {
        node{SAT}        
        child {
                node{SAT}
        		child {
                		node{UNSAT}
                		edge from parent
                		node[above] {$\eta_N$}
            	}
        		child {
                		node{\textbf{SAT}}
		    	edge from parent
                		node[above] {$\eta_P$}
            	}
            	child {
                		node{UNSAT}
                		edge from parent
                		node[above] {$\eta_U$}
            	}
	         child {
                		node{UNSAT}
                		edge from parent
                		node[above] {$\eta_O$}
            	}
                edge from parent
                node[above] {$\nu_E$}
            }
        child {
                node{SAT}
        		child {
                		node{UNSAT}
                		edge from parent
                		node[above] {$\eta_N$}
            	}
        		child {
                		node{\textbf{SAT}}
		    	edge from parent
                		node[above] {$\eta_P$}
            	}
            	child {
                		node{UNSAT}
                		edge from parent
                		node[above] {$\eta_U$}
            	}
	         child {
                		node{UNSAT}
                		edge from parent
                		node[above] {$\eta_O$}
            	}
                edge from parent
                node[above] {$\nu_O$}
            }
        edge from parent         
            node[above] {$\mu_P$}
    }
    child {
        node{SAT}        
        child {
                node{UNSAT}
                edge from parent
                node[above] {$\nu_E$}
            }
        child {
                node{SAT}
        		child {
                		node{\textbf{SAT}}
                		edge from parent
                		node[above] {$\eta_N$}
            	}
        		child {
                		node{\textbf{SAT}}
		    	edge from parent
                		node[above] {$\eta_P$}
            	}
            	child {
                		node{UNSAT}
                		edge from parent
                		node[above] {$\eta_U$}
            	}
	         child {
                		node{UNSAT}
                		edge from parent
                		node[above] {$\eta_O$}
            	}
                edge from parent
                node[above] {$\nu_O$}
            }
        edge from parent         
            node[above] {$\mu_O$}
    }
    child {
        node{SAT}        
        child {
                node{UNSAT}
                edge from parent
                node[above] {$\nu_E$}
            }
        child {
                node{SAT} 
        		child {
                		node{UNSAT}
                		edge from parent
                		node[above] {$\eta_P$}
            	}
        		child {
                		node{\textbf{SAT}}
		    	edge from parent
                		node[above] {$\eta_N$}
            	}
            	child {
                		node{\textbf{SAT}}
                		edge from parent
                		node[above] {$\eta_U$}
            	}
	         child {
                		node{UNSAT}
                		edge from parent
                		node[above] {$\eta_O$}
            	}
                edge from parent
                node[above] {$\nu_O$}
            }
        edge from parent         
            node[above] {$\mu_O$}
    };
\end{tikzpicture}
\caption{Incremental elimination of feasible modes}
\label{figure:sat_tree}
\end{figure}

\subsubsection{Deriving guards for the feasible mode combinations}\label{section:deriving-guards-sat}

For all feasible mode combinations, it is still necessary to compute (abstract) guards
which describe an over-approximation of those inputs that satisfy the respective mode. 
To illustrate the technique, consider the case where instruction $(1)$ underflows, instruction $(4)$
overflows and instruction $(6)$ is exact and non-negative. With $\varphi(\vec{X})$ encoding
the instructions that constitute the block as before, the formula $\xi(\vec{X})$ which encodes
this mode combination is thus defined as:
\[
\begin{array}{lll}
\xi(\vec{X}) & = & \varphi(\vec{X}) \wedge \mu_P(\vec{X}) \wedge \nu_E(\vec{X}) \wedge \eta_P(\vec{X})
\end{array}
\]
To derive an octagonal 
abstraction of the inputs that satisfy $\xi(\vec{X})$, first consider the problem of 
computing the least upper bound $d$ for the octagonal expression
$\signed{\vec{r0}} + \signed{\vec{r1}}$. To do so, let
$\kappa$ be a formula encoding $\signed{\vec{d}} = \signed{\vec{r0}} + \signed{\vec{r1}}$
where $\vec{d}$ is extended to $34$ bits to prevent wraps in
the octagonal expression (cp.~\cite[Sect.~3.3]{CKRW10}). Then, check
\[
\psi^1(\vec{X}) = \xi(\vec{X}) \wedge \kappa \wedge \neg \vec{d}[33]
\]
for satisfiability to derive a coarse approximation of $d$. The satisfiability of
$\psi^1(\vec{X})$ shows that $d \geq 0$. We thus proceed with testing
\[
\psi^2(\vec{X}) = \xi(\vec{X}) \wedge \kappa \wedge \neg \vec{d}[33] \wedge \vec{d}[32]
\]
for satisfiability. The unsatisfiability of $\psi^2(\vec{X})$ indicates $d < 2^{32}$.
Next we consider
\[
\psi^3(\vec{X}) = \xi(\vec{X}) \wedge \kappa \wedge \neg \vec{d}[33] \wedge \neg \vec{d}[32] \wedge \vec{d}[31]
\]
The unsatisfiability of $\psi^3(\vec{X})$ shows $d < 2^{31}$.
Then we test
\[
\psi^4(\vec{X}) = \xi(\vec{X}) \wedge \kappa \wedge \neg \vec{d}[33] \wedge \neg \vec{d}[32] \wedge \neg \vec{d}[31] \wedge \vec{d}[30]
\]
This and the ensuing formulae are all satisfiable.
The exact least upper bound is thus $\signed{\vec{d}} = 2^{30} + 2^{29} + \ldots + 2^{0} = 2^{31} - 1$ hence $\signed{\vec{r0}} + \signed{\vec{r1}} \leq 2^{31} - 1$. 

\begin{algorithm}[t!]
\caption{Compute the least signed value $d$ of the $k$-bit vector $\vec{d} = (\vec{d}[0], \ldots, \vec{d}[k-1])$ such that the Boolean formula $\varphi$ and
the inequality $\sum_{i=1}^n c_i \cdot \signed{\vec{v}_i} \leq \signed{\vec{d}}$ both hold; where the formula $\kappa$ encodes
$\sum_{i=1}^n c_i \cdot \signed{\vec{v}_i} = \signed{\vec{d}}$}
\label{algorithm:guards}
\begin{algorithmic}[1]
\REQUIRE {$\varphi$, $\kappa$}
\STATE {$\phi \gets \varphi \wedge \kappa$}
\STATE \COMMENT {check the sign}
\IF {$\phi \wedge \neg \vec{d}[k-1]$ is satisfiable}
	\STATE {$d \gets 0$}
	\STATE {$\phi \gets \phi \wedge \neg \vec{d}[k-1]$}
\ELSE
	\STATE {$d \gets -2^{k-1}$}
	\STATE {$\phi \gets \phi \wedge \vec{d}[k-1]$}
\ENDIF
\STATE \COMMENT {iterate over bits $k-2, \ldots 0$}
\FOR {$i = 1 \to k-1$}
	\IF {$\phi \wedge \vec{d}[j]$ is satisfiable}
		\STATE {$d \gets d + 2^{k-i-1}$}
		\STATE {$\phi \gets \phi \wedge \vec{d}[j]$}
	\ELSE
		\STATE {$\phi \gets \phi \wedge \neg \vec{d}[j]$}
	\ENDIF
	\STATE {$i \gets i+1$}
 \ENDFOR
 \RETURN {$d$}
\end{algorithmic}
\end{algorithm}

Alg.~\ref{algorithm:guards} presents this tactic for the general case of
maximising a linear expression of $n$ variables.
The algorithm relies on a propositional encoding
for an affine inequality constraint
$\sum_{i=0}^{n-1} c_i \cdot \signed{\vec{v}_i} \leq d$ 
where  $c_1, \ldots, c_n, d \in \mathbb{Q}$. To see that such
an encoding is possible assume, without loss of generality, that
the inequality is integral and $d$ is non-negative.
Then rewrite the inequality as
$\sum_{i=0}^{n-1} c^{+}_i \cdot \signed{\vec{v}_i} \leq d + \sum_{i=0}^{n-1} c^{-}_i \cdot \signed{\vec{v}_i}$
where 
$(c^{+}_1, \ldots, c^{+}_n), (c^{-}_1, \ldots, c^{-}_n) \in \mathbb{N}^n$
and $\mathbb{N} = \{ i \in \mathbb{Z} \mid 0 \leq i \}$.
Let $c^{+} = \sum_{i=0}^{n-1} c^{+}_i$
and
$c^{-} = \sum_{i=0}^{n-1} c^{-}_i$.
Since
$\signed{\vec{v}_i} \in [-2^{w-1}, 2^{w-1} - 1]$ for each bit-vector $\vec{v}_i$,
it follows that  computing the sums
$\sum_{i=1}^n c^{+}_i \cdot \signed{\vec{v}_i}$ and
$d + \sum_{i=1}^n c^{-}_i \cdot \signed{\vec{v}_i}$
with a signed
$1 + \lceil \log_2(1+\max(2^{w} \cdot c^{+}, b + 2^{w} \cdot c^{-})) \rceil$ bit
representation  is sufficient to avoid wraps~\cite[Sect.~3.2]{BK10}.
Lines 4--9 
provide special treatment for the sign. Lines 11--20 represent the core of the algorithm. 
Since the goal is maximisation, the algorithm instantiates each bit $\vec{d}$ with $1$, starting with $\vec{d}[k-2]$, and checks satisfiability
of the respective formula. If satisfiable, the bit $\vec{d}[k-i]$ is fixed at 1, and then the next
highest bit is examined.  If unsatisfiable, the bit $\vec{d}{[k-i]}$ can only take the value of 0, and
the algorithm moves on to maximise the next highest bit.  Variants of this algorithm have been reported elsewhere
\cite{BK10b,CLS08}.

Repeating 
this tactic for all five feasible modes, we compute the following
optimal octagonal guards:
\[
\begin{array}{rclc}
g_{O^{(1)},O^{(4)},U^{(6)}} &=  & 
\left\{ \begin{array}{ll}
2^{31} \leq \signed{\vec{r0}} + \signed{\vec{r1}} \leq 2^{31} & \wedge \\
1 \leq \signed{\vec{r0}} \leq 2^{31}-1 & \wedge \\
1 \leq \signed{\vec{r1}} \leq 2^{31}-1 \\
\end{array} \right. 
\\[2ex]
g_{O^{(1)},O^{(4)},N^{(6)}} & = & \left\{ \begin{array}{ll}
2^{31} + 1 \leq \signed{\vec{r0}} + \signed{\vec{r1}} \leq 2^{32} - 2 & \wedge \\
0 \leq \signed{\vec{r0}} \leq 2^{31}-1 & \wedge \\
0 \leq \signed{\vec{r1}} \leq 2^{31}-1 
\end{array} \right.
\\[2ex]
g_{U^{(1)},O^{(4)},P^{(6)}} & = &  
\left\{ \begin{array}{ll}
-2^{32} + 1 \leq \signed{\vec{r0}} + \signed{\vec{r1}} \leq -2^{31} - 1 & \wedge \\
-2^{31} \leq \signed{\vec{r0}} \leq -1 & \wedge \\
-2^{31} \leq \signed{\vec{r1}} \leq -1 \\
\end{array} \right. 
\\[2ex]
g_{U^{(1)},O^{(4)},N^{(6)}} & = &  
\left\{ \begin{array}{ll}
-2^{31} \leq \signed{\vec{r0}} \leq -2^{31} & \wedge \\
-2^{31} \leq \signed{\vec{r1}} \leq -2^{31} \\
\end{array} \right. 
\\[2ex]
g_{P^{(1)},E^{(4)},P^{(6)}} &=  & 
\left\{ \begin{array}{ll}
0 \leq \signed{\vec{r0}} + \signed{\vec{r1}} \leq 2^{31} - 1 & \wedge \\
0 \leq \signed{\vec{r1}} \leq 2^{31}-1
\end{array} \right. 
\\[2ex]
g_{P^{(1)},O^{(4)},P^{(6)}} & = & 
\left\{ \begin{array}{ll} 
0 \leq \signed{\vec{r0}} + \signed{\vec{r1}} \leq 2^{31}-1 & \wedge \\
-2^{31} \leq \signed{\vec{r1}} \leq -1 &
\end{array} \right. 
\\[2ex]
g_{N^{(1)},E^{(4)},N^{(6)}} &=  & 
\left\{ \begin{array}{ll}
-2^{31} +1 \leq \signed{\vec{r0}} + \signed{\vec{r1}} \leq - 1 & \wedge \\
-2^{31} \leq \signed{\vec{r1}} \leq -1 
\end{array} \right. 
\\[2ex]
g_{N^{(1)},O^{(4)},U^{(6)}} &=  & 
\left\{ \begin{array}{ll}
0 \leq \signed{\vec{r0}} \leq 0 & \wedge \\
-2^{31} \leq \signed{\vec{r1}} \leq -2^{31} \\
\end{array} \right. 
\\[2ex]
g_{N^{(1)},O^{(4)},N^{(6)}} & = & 
\left\{ \begin{array}{ll} 
-2^{31} +1 \leq \signed{\vec{r0}} + \signed{\vec{r1}} \leq -1 & \wedge \\
0 \leq \signed{\vec{r1}} \leq 2^{31} - 1
\end{array} \right. 
\\[2ex]
\end{array}
\]
Here, redundant inequalities, which are themselves entailed by the given guards, are omitted for clarity of presentation.
Note that if the non-negative and negative sub-cases where not distinguished then the feasible modes
$P^{(1)}, O^{(4)}, P^{(6)}$ and $N^{(1)}, O^{(4)}, N^{(6)}$ would be conflated into a single mode, for which the guard would be
$-2^{31} +1 \leq \signed{\vec{r0}} + \signed{\vec{r1}} \leq 2^{31} - 1 \wedge
-2^{31} \leq \signed{\vec{r1}} \leq 2^{31}  - 1$ which is almost vacuous.  The net effect of such a guard is that
its accompanying update operation would be applied frequently, possibly unnecessarily, inducing
a loss of precision.
This explains why it is attractive to separate the exact
modes into two sub-cases.  One can imagine enriching the modes by additionally
considering, for instance, the zero flag though as yet we have not encountered an example that warrants
resolving modes to this finer level of granularity. 

\subsubsection{Complexity} 

A total of $4 \cdot 34 + 4 \cdot 33$ SAT instances is solved for each octagonal guard.
This is due to the bit-extended representation for constraints $\pm v_1 \pm v_2 \leq d$, whereas 33 bits
are used for constraints $\pm v_1 \leq d$. While this may appear large, it is important to appreciate that
the number of SAT instances grows linearly with the bit-width. By way of comparison with~\cite{BK10},
adding a single propositional variable to a formula can increase the complexity of resolution
quadratically~\cite[Sect.~9.2.3]{KS08b}.

\subsection{Deriving template guards by range refinement}\label{section:guards-generalisation}

The generality of Alg.~\ref{algorithm:guards} hints that the approach to deriving guards
can be generalised to template inequalities where
the coefficients are restricted to take a finite range of possible values. Logahedra \cite{howe09logahedra} and octahedra \cite{CC04}
satisfy this property, the former being a class of two variable inequality where the coefficients
are limited a range $\{ -2^k, \ldots, -2^3, -2^2, -2^1, 0, 2^1, 2^2, 2^3, \ldots, 2^k \}$, and the latter being a class of
$n$ variable inequality where the coefficients are drawn from $\{ -1, 0, 1 \}$.   The approach straightforwardly
generalises to other finite classes of inequality, though it becomes less attractive
as the number of template inequalities increase.


\section{Deriving Updates with Affine Equations}
\label{section:updates}

Transformers over template constraints have been previously formulated
using quantification~\cite{BK10,Mon09}. To avoid this, we derive affine
relationships between output variables and input variables. These relations
are then lifted to symbolic constraints that detail how the bounds of an input
interval are mapped to the bounds of an output interval. The technique is then
refined to support octagons, so as to derive linear relationships between the
symbolic constants of the input octagon and the symbolic constants of the output octagon.
Note that Sect.~\ref{section:worked_examples:lifting}
and Sect.~\ref{section-medium} are just given for
pedagogical purposes; 
they build towards Sect.~\ref{section-exact}  which provides a linear symbolic update
operation that is optimal (if any equality relation exists between the input and output symbolic
constants then it will be found).  Sect.~\ref{section:worked_examples:lifting}
and Sect.~\ref{section-medium} motivate Sect.~\ref{section-exact} rather than provide
technical background, hence the latter section can be read independently of
the former sections if so desired.

\subsection{Inferring affine equalities}\label{sect-linear-affine}

Our algorithm computes an affine abstraction of the models for a
given mode-combination. To  solve for affine input-output relations, let 
$\vec{X}$ denote the set of bit-vectors as before. Consider the
Boolean formula $\xi(\vec{X})$ for the case where $(1)$ underflows,
$(4)$ overflows and $(6)$ is exact and non-negative. The process of deriving an affine
abstraction follows the scheme first presented in~\cite[Sect.~3.2]{BK10}. It
starts  with solving the formula $\xi(\vec{X})$, which produces a
model $\mathbf{m}_1$. Suppose the SAT solver yields:
\[
\mathbf{m}_1 = 
\left\{
\begin{array}{llll}
\signed{\vec{r0}'} = -2^{31} & \hspace{1.0em} & 
\signed{\vec{r1}'} = -1 \\
\signed{\vec{r0}} = -2^{31}+1 & \hspace{1.0em} & 
\signed{\vec{r1}} = -1
\end{array}
\right\}
\]
We can equivalently write $\mathbf{m}_1$ as an affine matrix, denoted $\mathbf{M}_1 \in \mathbb{Z}^{4 \times 5}$. With the
variable ordering $\langle \vec{r0}', \vec{r1'}, \vec{r0}, \vec{r1} \rangle$ on columns, this gives:
\[
\begin{array}{lll}
\mathbf{M}_1 & = & \left[
\begin{array}{cccc|l}
1 & 0 & 0 & 0 & -2^{31} \\
0 & 1 & 0 & 0 & -1 \\
0 & 0 & 1 & 0 & -2^{31}+1 \\
0 & 0 & 0 & 1 & -1
\end{array}
\right]
\end{array}
\]
We then add a disequality constraint $\signed{\vec{r1}} \neq -1$ to $\xi(\vec{X})$ in order to obtain a new solution that is not covered by $\mathbf{M}_1$. Denote this formula by $\xi'(\vec{X})$. Then, solving for $\xi'(\vec{X})$ produces a different model $\mathbf{m}_2$, say:
\[
\mathbf{m}_2 = 
\left\{
\begin{array}{lll}
\signed{\vec{r0}'} = -2^{31}+2 & \hspace{1.0em} & 
\signed{\vec{r1}'} = -3 \\
\signed{\vec{r0}} = -2^{31}+1 & \hspace{1.0em} &
\signed{\vec{r1}} = -3
\end{array}
\right\}
\]
Joining $\mathbf{M}_1$ with $\mathbf{M_2}$, which is likewise obtained from $\mathbf{m}_2$,
using the algorithm of M{\"u}ller-OIm and Seidl~\cite{MS04} yields a matrix that describes that
affine relations common to both models:
\[
\begin{array}{lll}
\mathbf{M}_1 \sqcup \mathbf{M}_2 & = & \left[
\begin{array}{cccc|l}
1 & 0 & 0 & 0 & -2^{31} \\
0 & 1 & 0 & 0 & -1 \\
0 & 0 & 1 & 0 & -2^{31}+1 \\
0 & 0 & 0 & 1 & -1
\end{array}
\right] 
\sqcup
\left[
\begin{array}{cccc|l}
1 & 0 & 0 & 0 & -2^{31}+2 \\
0 & 1 & 0 & 0 & -3 \\
0 & 0 & 1 & 0 & -2^{31}+1 \\
0 & 0 & 0 & 1 & -3
\end{array}
\right] \\
\\[-1ex]
& = &
\left[
\begin{array}{cccc|l}
1 & 1 & 0 & 0 & -2^{31}-1 \\
0 & 1 & 0 & -1 & 0 \\
0 & 0 & 1 & 0 & -2^{31}+1
\end{array}
\right]
\end{array}
\]
Our algorithm now attempts to find a model that violates the constraint given through the last row, that is, $\signed{\vec{r0}} = -2^{31}+1$. Adding a disequality constraint to $\xi'(\vec{X})$ yields a new formula $\xi''(\vec{X})$, for which a SAT solver finds a model:
\[
\mathbf{m}_3 = 
\left\{
\begin{array}{lll}
\signed{\vec{r0}'} = -2^{31} & \hspace{1.0em} & \signed{\vec{r1}'} = -4 \\
\signed{\vec{r0}} = -2^{31}+4 & \hspace{1.0em} & \signed{\vec{r1}} = -4
\end{array}
\right\}
\]
Then, we join $\mathbf{M}_1 \sqcup \mathbf{M}_2$ with $\mathbf{M}_3$ to give:
\[
\begin{array}{lllll}
(\mathbf{M}_1 \sqcup \mathbf{M}_2) \sqcup \mathbf{M}_3 & = &
\left[
\begin{array}{cccc|l}
1 & 1 & 0 & 0 & -2^{31}-1 \\
0 & 1 & 0 & -1 & 0 \\
0 & 0 & 1 & 0 & -2^{31}+1
\end{array}
\right]
\sqcup
\left[
\begin{array}{cccc|l}
1 & 0 & 0 & 0 & -2^{31} \\
0 & 1 & 0 & 0 & -4 \\
0 & 0 & 1 & 0 & -2^{31}+4 \\
0 & 0 & 0 & 1 & -4
\end{array}
\right] \\
& = &
\left[
\begin{array}{cccc|l}
1 & 0 & 1 & 1 & -2^{32} \\
0 & 1 & 0 & -1 & 0
\end{array}
\right]
\end{array}
\]
Adding a disequality constraint to suppress $\signed{\vec{r1}'} - \signed{\vec{r1}} = 0$ yields an unsatisfiable formula, likewise for $\signed{\vec{r0}'} + \signed{\vec{r1}} + \signed{\vec{r0}} = -2^{32}$. Indeed, we have
\[
\begin{array}{lll}
(\mathbf{M}_1 \sqcup \mathbf{M}_2) \sqcup \mathbf{M}_3 & = & \bigsqcup_{i \in \mathbb{N}} \mathbf{M}_i
\end{array}
\]
where $\mathbf{M}_i$ are matrices describing different 
models $\vec{m}_i$ of $\xi(\vec{X})$. Indeed, an affine
summary of a mode-combination is in some sense universally 
quantified, since its relation is satisfied by every model. Moreover
$(\mathbf{M}_1 \sqcup \mathbf{M}_2) \sqcup \mathbf{M}_3$ 
represents the best affine abstraction of $\xi(\vec{X})$ \cite{BK10,KS10}. 
Note too that the chain-length in the affine domain is linear in the number of variables in the system~\cite{Kar76}. Thus, the number of iterations required to compute a
fixed point is bounded by the number of variables and does not depend on the bit-width. 

The resulting equations, however, express relationships
between variables but not between the ranges of the input and output intervals. As it turns 
out, we can lift $(\mathbf{M}_1 \sqcup \mathbf{M}_2) \sqcup \mathbf{M}_3$ 
to an equation system over intervals by applying a set of straightforward 
transformations.  This is arguably the most natural way of deriving a transformer for intervals, though
we shall see that it does not extend well to octagons.


%
%

\subsection{Lifting affine equalities to interval updates}
\label{section:worked_examples:lifting}

We explain how to transform the resulting affine system
$(\mathbf{M}_1 \sqcup \mathbf{M}_2) \sqcup \mathbf{M}_3$
over variables in $\vec{X}$ into an equation system over
range boundaries that prescribes an update. To do so, let $\vec{V} \subseteq \vec{X}$
denote the bit-vectors on entry of the block, and likewise let
$\vec{V}' \subseteq \vec{X}$ denote the bit-vectors on exit.
Further, introduce sets of fresh variables
\[
\begin{array}{lll}
\vec{V}_{\mymin} = \{ \vec{r0}_{\mymin}, \vec{r1}_{\mymin} \} & \hspace{1em} &
\vec{V}_{\mymax} = \{ \vec{r0}_{\mymax}, \vec{r1}_{\mymax} \} \\
\vec{V}'_{\mymin} = \{ \vec{r0}'_{\mymin}, \vec{r1}'_{\mymin} \} & \hspace{1em} &
\vec{V}'_{\mymax} = \{ \vec{r0}'_{\mymax}, \vec{r1}'_{\mymax} \}
\end{array}
\]
to represent symbolic boundaries of each bit-vector in $\vec{V} \cup \vec{V}'$.
If necessary transform the equations such that the left-hand side consists of
only one variable in $\vec{V}'$. For the above system, this transformation gives:
\[
\begin{array}{lll}
\signed{\vec{r1}'} & = & \signed{\vec{r1}} \\
\signed{\vec{r0}'} & = & -\signed{\vec{r0}} - \signed{\vec{r1}} - 2^{32}
\end{array}
\]
These equations imply the following affine relations on interval boundaries:
\[
\begin{array}{lllllll}
\signed{\vec{r1}'}_{\mymax} & = & \signed{\vec{r1}'}_{\mymax} & 
\hspace{2em} & 
\signed{\vec{r0}'}_{\mymax} & = & -\signed{\vec{r1}}_{\mymin} - \signed{\vec{r0}}_{\mymin}  - 2^{32} \\
\signed{\vec{r1}'}_{\mymin} & = & \signed{\vec{r1}'}_{\mymin}
& &
\signed{\vec{r0}'}_{\mymin} & = & -\signed{\vec{r1}}_{\mymax} - \signed{\vec{r0}}_{\mymax} - 2^{32}
\end{array}
\]
To derive such as system, transform each of the original equations into the form
\[
\begin{array}{lll}
\lambda_{\vec{v}'} \cdot \vec{v}' & = & \sum_{\vec{v} \in \vec{V}} \lambda_{\vec{v}} \cdot \vec{v} + d
\end{array}
\]
where $\vec{v}' \in \vec{V}'$, $\lambda_{\vec{v}'} > 0$ and $\lambda_{\vec{v}} \in \mathbb{Z}$ for all $\vec{v} \in \vec{V}$. This can always be achieved due to the variable ordering. For example, the system below on the left can be transformed into the system on the right
by applying elementary row operations:
\[
\begin{array}{lll}
\left[
\begin{array}{rrrr|l}
1 & -1 & 0 & 0 & 1 \\
0 & 1 & 0 & -1 & 2
\end{array}
\right]
& \hspace{1em} \rightsquigarrow \hspace{1em} &
\left[
\begin{array}{rrrr|l}
1 & 0 & 0 & -1 & 3 \\
0 & 1 & 0 & -1 & 2
\end{array}
\right]
\end{array}
\]
Note that the leading coefficients are positive. We then replace each original equation
by a pair of equations as follows:
\[
\begin{array}{lll}
\lambda_{\vec{v}'} \cdot \vec{v}'_{\mymax} & = & \sum_{\vec{v} \in \vec{X}} \lambda_{\vec{v}} \cdot \beta(\lambda_{\vec{v}}, \vec{v}) + d \\
\lambda_{\vec{v}'} \cdot \vec{v}'_{\mymin} & = & \sum_{\vec{v} \in \vec{X}} \lambda_{\vec{v}} \cdot \beta(-\lambda_{\vec{v}}, \vec{v}) + d
\end{array}
\]
where the map $\beta : \mathbb{Z} \times \vec{V} \rightarrow (\vec{V}_{\mymin} \cup \vec{V}_{\mymax})$
is defined as:
\[
\begin{array}{lll}
\beta(\lambda,\vec{v}) & = &
	\left\{
	\begin{array}{ll}
		\vec{v}_{\mymin} & : \text{if } \lambda < 0 \\
	 	\vec{v}_{\mymax} & : \text{otherwise}
	\end{array}
	\right.
\end{array}
\]
The key idea when constructing the upper bound is to replace each occurrence
of a variable in the original system with its upper bound in case its coefficient is
positive, and with its lower bound otherwise. This task is performed by $\beta$. 
An analogous technique is applied when defining the lower bound. Applying this
technique to all affine systems, we obtain the following five transfer functions over
symbolic ranges, rather than concrete variables
(with the identity constraints on $\vec{r1}'_{\mymin}$ and $\vec{r1}'_{\mymax}$ omitted):
\[
\begin{array}{llllllllrllllll}
f_{O^{(1)},O^{(4)},U^{(6)}} & = &
\left\{
\begin{array}{llll}
(\signed{\vec{r0}_{\mymin}'} & = & -2^{31}) & \wedge \\ 
 (\signed{\vec{r0}_{\mymax}'} & = & -2^{31})
\end{array}\right.
\\[2ex]
f_{O^{(1)},O^{(4)},N^{(6)}} & = &
\left\{
\begin{array}{llll}
(\signed{\vec{r0}_{\mymin}'} & = & 2^{32} - \signed{\vec{r0}_{\mymax}} - \signed{\vec{r1}_{\mymax}}) & \wedge \\ 
 (\signed{\vec{r0}_{\mymax}'} & = & 2^{32} - \signed{\vec{r0}_{\mymin}} - \signed{\vec{r1}_{\mymin}})
\end{array}\right.
\\[2ex]
f_{U^{(1)},O^{(4)},P^{(6)}} & = &
\left\{
\begin{array}{llll}
(\signed{\vec{r0}_{\mymin}'} & = & -2^{32} - \signed{\vec{r0}_{\mymax}} - \signed{\vec{r1}_{\mymax}}) & \wedge \\ 
 (\signed{\vec{r0}_{\mymax}'} & = & -2^{32} - \signed{\vec{r0}_{\mymin}} - \signed{\vec{r1}_{\mymin}})
\end{array}\right.
\\[2ex]
f_{U^{(1)},O^{(4)},N^{(6)}} & = &
\left\{
\begin{array}{llll}
(\signed{\vec{r0}_{\mymin}'} & = & 0) & \wedge \\ 
 (\signed{\vec{r0}_{\mymax}'} & = & 0)
\end{array}\right.
\\[2ex]
f_{P^{(1)},E^{(4)},P^{(6)}} & = &
\left\{
\begin{array}{llll}
(\signed{\vec{r0}_{\mymin}'} & = & \signed{\vec{r0}_{\mymin}} + \signed{\vec{r1}_{\mymin}}) & \wedge \\ 
 (\signed{\vec{r0}_{\mymax}'} & = & \signed{\vec{r0}_{\mymax}} + \signed{\vec{r1}_{\mymax}})
\end{array}\right.
\\[2ex]
f_{P^{(1)},O^{(4)},P^{(6)}} & = &
\left\{
\begin{array}{llll}
(\signed{\vec{r0}_{\mymin}'} & = & -\signed{\vec{r0}_{\mymax}} - \signed{\vec{r1}_{\mymax}}) & \wedge \\ 
 (\signed{\vec{r0}_{\mymax}'} & = & -\signed{\vec{r0}_{\mymin}} - \signed{\vec{r1}_{\mymin}})
\end{array}\right.
\\[2ex]
f_{N^{(1)},E^{(4)},N^{(6)}} & = &
\left\{
\begin{array}{llll}
(\signed{\vec{r0}_{\mymin}'} & = & \signed{\vec{r0}_{\mymin}} + \signed{\vec{r1}_{\mymin}}) & \wedge \\ 
 (\signed{\vec{r0}_{\mymax}'} & = & \signed{\vec{r0}_{\mymax}} + \signed{\vec{r1}_{\mymax}})
\end{array}\right.
\\[2ex]
f_{N^{(1)},O^{(4)},U^{(6)}} & = &
\left\{
\begin{array}{llll}
(\signed{\vec{r0}_{\mymin}'} & = & -2^{31}) & \wedge \\ 
 (\signed{\vec{r0}_{\mymax}'} & = & -2^{31})
\end{array}\right.
\\[2ex]
f_{N^{(1)},O^{(4)},N^{(6)}} & = &
\left\{
\begin{array}{llll}
(\signed{\vec{r0}_{\mymin}'} & = & -\signed{\vec{r0}_{\mymax}} - \signed{\vec{r1}_{\mymax}}) & \wedge \\ 
 (\signed{\vec{r0}_{\mymax}'} & = & -\signed{\vec{r0}_{\mymin}} - \signed{\vec{r1}_{\mymin}})
\end{array}\right.
\\[2ex]
\end{array}
\]
To illustrate the accuracy of this result, consider the application of the transfer function $f_{U^{(1)},O^{(4)},P^{(6)}}$ to the input intervals defined by:
\[
\begin{array}{@{}l@{\qquad}l@{\qquad}l@{\qquad}l@{}}
\signed{\vec{r0}_{\mymin}} = -2^{31} + 1 &
\signed{\vec{r0}_{\mymax}} = -2^{31}+4 &
\signed{\vec{r1}_{\mymin}} = -20 & 
\signed{\vec{r1}_{\mymax}} = -10
\end{array}
\]
Then, the above transfer function defines the output intervals by modelling the wrap that occurs in the
first instruction \mbox{\texttt{ADD R0 R1}} to give $\signed{\vec{r0}'_{\mymin}} = -2^{31}+6$ and
$\signed{\vec{r0}'_{\mymax}} = -2^{31}+19$. 
When multiple guards are applicable, however,
a merge operation need be applied to combine the results of the different updates. This looses
information. Further details of the evaluation mechanism are
discussed in section~\ref{section:evaluating}.  

It is interesting to compare this with how an interval analysis would proceed for the block which, recall,
is listed in Fig.~\ref{figure:example_program}.
Initially, the \texttt{R0}, \texttt{R1} and \texttt{R2} would respectively be assigned the intervals
$[-2^{31} + 1, -2^{31} + 4]$, $[-20, -10]$ and $[-2^{31}, 2^{31} - 1]$
the third interval being vacuous. The 
\mbox{\texttt{ADD R0 R1}} instruction will assign \texttt{R0} to $[2^{31} - 19, 2^{31} - 6]$ simulating
an underflow
and  \mbox{\texttt{MOV R2 R0}} will update \texttt{R2} to $[2^{31} - 19, 2^{31} - 6]$.
The \mbox{\texttt{EOR R2 R1}} instruction will then reassign \texttt{R2} to $[-2^{31}, -2^{31} + 2^{29} - 1]$
which is adjusted  to $[0, 2^{30} - 1]$ by \mbox{\texttt{LSL R2}}. In a carefully constructed interval
analysis the transfer function for \mbox{\texttt{LSR R2}} will also assign the carry flag to 1. In such an analysis, the instruction
\mbox{\texttt{SBC R2 R2}} might even assign \texttt{R2} to [-1, -1] rather than a wider interval.  Under this assumption
\mbox{\texttt{ADD R0 R2}} will update \texttt{R0} to $[2^{31} - 20, 2^{31} - 7]$.  Then, since the sign bit is clear and
following 26
high bits of \texttt{R0} are set for all values in the interval $[2^{31} - 20, 2^{31} - 7]$, a transfer function
for \mbox{\texttt{EOR R0 R2}} could conceivable assign \texttt{R0} to $[-2^{31}, -2^{31}+31]$.

\subsection{Lifting affine equalities to octagonal updates}
\label{section-medium}

Consider now the more general problem of deriving a transfer function for octagons for \texttt{ADD R0 R1; LSL R0} where \texttt{ADD} and \texttt{LSL} operate in exact non-negative modes. Computing the affine relation for this mode-combination gives $(\signed{\vec{r0}'} = 2 \cdot \signed{\vec{r0}} + 2 \cdot \signed{\vec{r1}}) \wedge (\signed{\vec{r1}'} = \signed{\vec{r1}})$. We aim to construct an update that maps octagonal input constraints with symbolic constants to octagonal outputs likewise with symbolic constants of the form:
\[
\begin{array}{lll}
\left\{
\begin{array}{rllllll}
\signed{\vec{r0}} & \leq & d_1 \\
\signed{\vec{r1}} & \leq & d_2 \\
-\signed{\vec{r0}} & \leq & d_3 \\
-\signed{\vec{r1}} & \leq & d_4 \\
\hline
\signed{\vec{r0}} + \signed{\vec{r1}} & \leq & d_5 \\
-\signed{\vec{r0}} - \signed{\vec{r1}} & \leq & d_6 \\
-\signed{\vec{r0}} + \signed{\vec{r1}} & \leq & d_7 \\
\signed{\vec{r0}} - \signed{\vec{r1}} & \leq & d_8 \\
\end{array}
\right\}
&
\hspace{1em}
\leadsto
\hspace{1em}
&
\left\{
\begin{array}{rllllll}
\signed{\vec{r0}'} & \leq & 2 \cdot (d_1 + d_2) \\
\signed{\vec{r1}'} & \leq & d_2 \\
-\signed{\vec{r0}'} & \leq & 2 \cdot (d_3 + d_4) \\
-\signed{\vec{r1}'} & \leq & d_4 \\
\signed{\vec{r0}'} + \signed{\vec{r1}'} & \leq & 2 \cdot d_1 + 3 \cdot d_2 \\
-\signed{\vec{r0}'} - \signed{\vec{r1}'} & \leq & 2 \cdot d_3 + 3 \cdot d_4 \\
-\signed{\vec{r0}'} + \signed{\vec{r1}'} & \leq & 2 \cdot (d_3 + d_4) + d_2  \\
\signed{\vec{r0}'} - \signed{\vec{r1}'} & \leq & 2 \cdot (d_1 + d_2) + d_4 \\
\end{array}
\right\}
\end{array}
\]
We start by constructing an update operation that uses
the unary input constraints only, which appear above the bar
separator. We modify the method presented in
Sect.~\ref{section:worked_examples:lifting} so as to express output constraints in terms of
symbolic variables $d_1, \dots, d_4$ from
the input constraints. We obtain the four output unary constraints by an
analogous technique as before by substituting the symbolic minima and maxima for the symbolic output constants. The binary output
constraints are derived by linear combinations of the unary output constraints. 
Since the output constraints do not use relational information from the inputs,
such as $\signed{\vec{r0}} + \signed{\vec{r1}} \leq d_5$,
we obtain a sub-optimal update. To illustrate, suppose $0 \leq \signed{\vec{r0}} \leq 4$, $0 \leq \signed{\vec{r1}} \leq 1$ and $\signed{\vec{r0}} + \signed{\vec{r1}} \leq 4$. Then we derive:
\[
\begin{array}{lllll}
0 \leq \signed{\vec{r0}'} \leq 10 & \hspace{1em} &
0 \leq \signed{\vec{r1}'} \leq 1 & \hspace{1em} &
0 \leq \signed{\vec{r0}'} + \signed{\vec{r1}'} \leq 11  
\end{array}
\]
An optimal transfer function, however, would derive $\signed{\vec{r0}'} \leq 8$ and $\signed{\vec{r0}'} + \signed{\vec{r1}'} \leq 8$. Although the above method fails to propagate the effect of some inputs into the outputs, it retains the property that the update can be constructed straightforwardly by lifting the affine relations. In what follows, we will describe how to derive more precise affine relations for the outputs.

\subsection{Inferring affine inequalities for octagonal updates}
\label{section-exact}

To derive more precise affine updates for octagons, let $\xi(\vec{X})$ denote the
propositional encoding for
\mbox{\tt ADD R0 R1; LSL R0}
where again \mbox{\tt ADD} and \mbox{\tt LSL}
operate in exact non-negative modes.
Consider inequality
$\signed{\vec{r0}'} \leq d'_1$ in the output octagon and in particular
the problem of discovering
a relationship between $d'_1$ and the symbolic constants
$d_1, \ldots, d_8$ of the input octagon, as detailed previously.

\begin{table}[!t]
\caption{Intermediate results for inferring exact affine transformers for octagons}
\label{table:affine_octagons}
\[
\begin{array}{l|r|rrrrrrrr|r}
    & \signed{\vec{d}'_1} & \signed{\vec{d}_1} & \signed{\vec{d}_2} & \signed{\vec{d}_3} & \signed{\vec{d}_4} &
	                  \signed{\vec{d}_5} & \signed{\vec{d}_6} & \signed{\vec{d}_7} & \signed{\vec{d}_8} & \max(\signed{\vec{d}'}) \\ \hline
\vec{m}_1 & 1  &   1 &   1 &   0 &   0 &   1 &   0 &   1 &   1  & 2 \\  
\vec{m}_2 & 8  &   3 &   3 &  -1 &  -1 &   5 &  -2 &   2 &   0  & 10 \\
\vec{m}_3 & 22 &   8 &   7 &   0 &   1 &  13 &   3 &   4 &   0  & 26 \\
\vec{m}_4 & 4  &   0 &   3 &   2 &   0 &   3 &   1 &   6 &   3  &  6
\end{array}
\]
\end{table}

We proceed by introducing signed 34-bit vectors
$\vec{d}_1, \ldots, \vec{d}_8$ to represent the symbolic constants $d_1, \ldots, d_8$.
Further, let $\kappa$ denote a Boolean formula that holds
iff the eight inequalities
$\signed{\vec{r0}} \leq \signed{\vec{d_1}}$, \ldots,
$\signed{\vec{r0}} - \signed{\vec{r1}} \leq \signed{\vec{d_8}}$ simultaneously hold.
Furthermore, let $\eta$ denote a formula
that encodes the equality $\signed{\vec{r0}'} = \signed{\vec{d'_1}}$ 
where $\vec{d'_1}$ is a signed bit-vector representing $d'_1$.
Presenting the compound formula $\kappa \wedge \xi(\vec{X}) \wedge \eta$ to a SAT solver
produces a model:
\[
\vec{m}_1 = 
\left\{
\begin{array}{llllllllllllllll}
\signed{\vec{d'_1}} = 1, &
\signed{\vec{d_1}} = 1, &
\signed{\vec{d_2}} = 1, &
\ldots,
\signed{\vec{d_7}} = 1, &
\signed{\vec{d_8}} = 1 &
\end{array}
\right\}
\]
which is fully detailed in Tab.~\ref{table:affine_octagons}. The assignment 
$\signed{\vec{d'_1}} = 1$ does not necessarily represent the maximum value of 
$\signed{\vec{d'_1}}$ for the partial assignment
$\signed{\vec{d_1}} = 1$,
\ldots,
$\signed{\vec{d_8}} = 1$. 
Thus let $\zeta_1$ denote a formula that holds
iff
$\signed{\vec{d_1}} = 1$,
\ldots,
$\signed{\vec{d_8}} = 1$ all hold.
Then range refinement can be applied to find the maximal
value of
$\signed{\vec{d'_1}}$ subject to
$\kappa \wedge \xi(\vec{X}) \wedge \eta \wedge \zeta$.  This gives
$\signed{\vec{d'_1}} = 2$ and
a model:
\[
\begin{array}{lll}
\vec{m}'_1 & = & \left\{ \signed{\vec{d'_1}} = 2, \signed{\vec{d_1}} = 1, \ldots, \signed{\vec{d_8}} = 1 \right\}
\end{array}
\]
An affine summary of all such maximal models can be found by interleaving
range refinement with affine join.  Thus suppose the matrix $\mathbf{M}_1$ is
constructed from $\vec{m}'_1$ by using the variable ordering $\langle d'_1, d_1, \ldots, d_8 \rangle$ on
columns:
\[
\mathbf{M}_1 = \left[
\begin{array}{@{}ccccccccc|l@{}}
1 & 0 & 0 & 0 & 0 & 0 & 0 & 0 & 0 & 2 \\
0 & 1 & 0 & 0 & 0 & 0 & 0 & 0 & 0 & 1 \\
0 & 0 & 1 & 0 & 0 & 0 & 0 & 0 & 0 & 1 \\
0 & 0 & 0 & 1 & 0 & 0 & 0 & 0 & 0 & 0 \\
0 & 0 & 0 & 0 & 1 & 0 & 0 & 0 & 0 & 0 \\
0 & 0 & 0 & 0 & 0 & 1 & 0 & 0 & 0 & 1 \\
0 & 0 & 0 & 0 & 0 & 0 & 1 & 0 & 0 & 0 \\
0 & 0 & 0 & 0 & 0 & 0 & 0 & 1 & 0 & 1 \\
0 & 0 & 0 & 0 & 0 & 0 & 0 & 0 & 1 & 1 \\
\end{array}
\right]
\]
The method proceeds in an analogous fashion to before by constructing
a formula $\mu$ that holds iff  $\signed{\vec{d_8}} \neq 1$ holds.  Solving the formula
$\kappa \wedge \xi(\vec{X}) \wedge \eta \wedge \mu$ gives the model $\vec{m}_2$
detailed in Tab.~\ref{table:affine_octagons}. The model
$\vec{m}_2$, itself, defines a formula $\zeta_2$
that is equi-satisfiable with the conjunction of
$\signed{\vec{d_1}} = 3$,
\ldots,
$\signed{\vec{d_8}} = 0$. Maximising 
$\signed{\vec{d}'_1}$ subject to
$\kappa \wedge \xi(\vec{X}) \wedge \eta \wedge \zeta_2$ gives
$\signed{\vec{d}'_1} = 10$ which defines
the model
\[
\begin{array}{lll}
\vec{m}'_2 & = & \left\{ \signed{\vec{d'_1}} = 10, \signed{\vec{d_1}} = 3, \ldots, \signed{\vec{d_8}} = 0 \right\}
\end{array}
\]
and $\mathbf{M}_2$, which in turn yields the join $\mathbf{M}_1 \sqcup \mathbf{M}_2$ as follows:
\[
\begin{array}{rl}
\mathbf{M}_1 \sqcup \mathbf{M}_2 = &
\left[
\begin{array}{@{}ccccccccc|l@{}}
1 & 0 & 0 & 0 & 0 & -2 & 0 & 0 & 0 & 0 \\
0 & 1 & -1 & 0 & 0 & 0 & 0 & 0 & 0 & 0 \\
0 & 0 & 0 & 1 & -2 & 0 & 0 & 0 & 0 & 0 \\
0 & 0 & 0 & 0 & 0 & 0 & 1 & 2 & 0 & 2 \\
0 & 0 & 0 & 0 & 0 & 0 & 0 & 1 & 1 & 1 \\
\end{array}
\right]
\end{array}
\]
Repeating this process two more times then gives:
\[
\begin{array}{lll}
\vec{m}'_3 & = & \left\{ \signed{\vec{d'_1}} = 26, \signed{\vec{d_1}} = 8, \ldots, \signed{\vec{d_8}} = 0 \right\} \\[0.2ex]
\vec{m}'_4 & = & \left\{ \signed{\vec{d'_1}} = 6, \signed{\vec{d_1}} = 0, \ldots, \signed{\vec{d_8}} = 3 \right\}
\end{array}
\]
\[
\begin{array}{rl}
\mathbf{M}_1 \sqcup \mathbf{M}_2 \sqcup \mathbf{M}_3 = &
\left[
\begin{array}{@{}ccccccccc|l@{}}
1 & 0 & 0 & 0 & 0 & -2 & 0 & 0 & 0 & 0 \\
0 & 1 & -1 & 1 & -1 & 0 & 0 & 0 & 0 & 0 \\
\end{array}
\right]
\\[2.5ex]
\mathbf{M}_1 \sqcup \mathbf{M}_2 \sqcup \mathbf{M}_3 \sqcup \mathbf{M}_4 = &
\left[
\begin{array}{@{}ccccccccc|l@{}}
1 & 0 & 0 & 0 & 0 & -2 & 0 & 0 & 0 & 0 \\
\end{array}
\right]
\end{array}
\]
The system $\mathbf{M}_1 \sqcup \mathbf{M}_2 \sqcup \mathbf{M}_3 \sqcup \mathbf{M}_4$
then expresses the relationship
$\signed{\vec{d}'_1} = 2 \cdot \signed{\vec{d}_5}$.  In summary, each iteration of the algorithm
involves the following steps: find a model of $\kappa \wedge \xi(\vec{X}) \wedge \eta \wedge \mu$ where
$\mu$ ensures that the model is not already summarised by $\sqcup_{i=1}^{\ell} \mathbf{M}_i$;
apply range refinement to maximise $\signed{ \vec{d}'_1} $
whilst keeping $\signed{ \vec{d}'_1} $, $\signed{ \vec{d}_1} $, \ldots, $\signed{ \vec{d}_8} $ fixed; 
join the resulting model with $\sqcup_{i=1}^{\ell} \mathbf{M}_i$ to give $\sqcup_{i=1}^{\ell+1} \mathbf{M}_i$.

To verify that $\signed{\vec{d}'_1} = 2 \cdot \signed{\vec{d}_5}$ is a fixed point, unlike before, it not sufficient
to impose the disequality $\signed{\vec{d}'_1} \neq 2 \cdot \signed{\vec{d}_5}$ and check for unsatisfiability.
This is because $\signed{\vec{d}'_1}$ is defined through maximisation. Instead the check amounts to testing whether
$\kappa \wedge \xi(\vec{X}) \wedge \eta$ is unsatisfiable when combined with a formula encoding the strict inequality
$\signed{\vec{d}'_1} > 2 \cdot \signed{\vec{d}_5}$
(note that if $\signed{\vec{d}'_1} > 2 \cdot \signed{\vec{d}_5}$ holds
then it follows that $\signed{\vec{d}'_1} \neq 2 \cdot \signed{\vec{d}_5}$ holds).
Since the combined system is
unsatisfiable, we conclude that the update for this mode-combination
includes $d'_1 = 2 \cdot d_5$. The complete affine update consists of:
\[
\begin{array}{l@{\qquad\qquad}l}
\begin{array}{rcl}
d'_1 & = & 2 \cdot d_5 \\ 
d'_2 & = & d_2 \\ 
d'_3 & = & 2 \cdot d_6 \\ 
d'_4 & = & d_4 
\end{array}
&
\begin{array}{rcl}
d'_5 & = & 2 \cdot d_5 + d_2 \\ 
d'_6 & = & 2 \cdot d_6 + d_4 \\ 
d'_7 & = & 2 \cdot d_6 + d_2 \\ 
d'_8 & = & 2 \cdot d_5 + d_4
\end{array}
\end{array}
\]
This result is superior to that computed in Sect.~\ref{section-medium}. To illustrate, consider
again an input octagon defined by $0 \leq \signed{\vec{r0}} \leq 4$, $0 \leq \signed{\vec{r1}} \leq 1$
and $\signed{\vec{r0}} + \signed{\vec{r1}} \leq 4$, hence:
\[
\begin{array}{llll}
d_1 = 4 & \hspace{2em} & d_3 = 0 \\
d_2 = 1 & & d_4 = 0 \\
d_5 = 4
\end{array}
\] 
Applying the computed transformer to derive $d'_5$ on output gives:
\[
\begin{array}{lllll}
d'_5 & = & 2 \cdot 4 + 1 & = & 9
\end{array}
\]
Hence, we have $\signed{\vec{r0}'} + \signed{\vec{r1}'} \leq 9$,
whereas the previously discussed technique based on applying
the $\beta$ map yields  $\signed{\vec{r0}'} + \signed{\vec{r1}'} \leq 11$.
Indeed, these linear symbolic update operations are optimal in the sense that
if a symbolic output constant $d'_j$ is equal to a linear function of the symbolic
input constants $d_1, \ldots, d_8$, then that function will be derived.

Interestingly, Min\'{e} \cite[Fig.~27]{Min06} also discusses the relative
precision of transfer functions, though where the
base semantics is polyhedral rather than Boolean. Using his
classification, the transfer functions
derived using the synthesis techniques presented in Sect.~\ref{section-medium} 
and Sect.~\ref{section-exact} might be described as
medium and exact.  The following theorem confirms this intuition.  For ease
of presentation, the result states the exactitude of the update on
the symbolic constant $d'_1$; analogous results hold for updates on
$d'_2, \ldots, d'_8$.

\begin{thm} \rm Suppose an octagonal update is derived of the form
$\mathbf{M} \langle d'_1, d_1, \ldots, d_8, -1 \rangle = 0$. Moreover suppose that
\begin{iteMize}{$\bullet$}

\item
for all values of
$\signed{\vec{r0}}, \signed{\vec{r1}}$ such that
\mbox{$\signed{\vec{r0}} \leq d_1$}, \ldots,
\mbox{$\signed{\vec{r0}} - \signed{\vec{r1}} \leq d_8$} 
and $\xi(\vec{X})$ hold it follows that
\mbox{$\signed{\vec{r0}'} \leq c +  c_1 \cdot d_1 + \ldots + c_8 \cdot d_8$} holds

\item
for all values of
\mbox{$\signed{\vec{r0}}, \signed{\vec{r1}}$} 
there exists a value of
$\signed{\vec{r0}'}$ 
such that
\mbox{$\signed{\vec{r0}} \leq d_1$}, \ldots,
\mbox{$\signed{\vec{r0}} - \signed{\vec{r1}} \leq d_8$}, 
$\xi(\vec{X})$ 
and
\mbox{$\signed{\vec{r0}'} = c +  c_1 \cdot d_1 + \ldots + c_8 \cdot d_8$} hold

\end{iteMize} 
Then $\mathbf{M} \langle d'_1, d_1, \ldots, d_8, -1 \rangle = 0 \models d'_1 = c +  c_1 \cdot d_1 + \ldots + c_8 \cdot d_8$
\end{thm}

\proof  
Suppose that $\mathbf{M}$ is derived by $\mathbf{M} = \mathbf{M}_1 \sqcup \mathbf{M}_2 \sqcup \ldots \sqcup \mathbf{M}_\ell$.
Suppose $\mathbf{M}_1$ is constructed from the model $\mathbf{m}_1 = \{ d'_1 = v'_1, d_1 = v_1, \ldots, d_8 = v_8 \}$
where the value  $v'_1$ is maximal. Yet $\mathbf{m}_1$ is derived from a formula
that encodes the equality $\signed{\vec{r0}'} = \signed{\vec{d'_1}}$ 
where $\vec{d'_1}$ is a signed bit-vector representing $d'_1$.  Since $v'_1$ is maximal
it follows that the value of $\signed{\vec{r0}'}$ is maximal hence
$(d'_1 = v'_1) \wedge (d_1 = v_1) \wedge \ldots \wedge (d_8 = v_8) \models d'_1 = c +  c_1 \cdot d_1 + \ldots + c_8 \cdot d_8$
by the two assumptions.  Therefore 
$\mathbf{M}_1 \langle d'_1, d_1, \ldots, d_8, -1 \rangle = 0 \models d'_1 = c +  c_1 \cdot d_1 + \ldots + c_8 \cdot d_8$
since $\mathbf{M}_1 = [I \mid \langle v'_1, v_1, \ldots, v_8 \rangle]$. Likewise 
$\mathbf{M}_i \langle d'_1, d_1, \ldots, d_8, -1 \rangle = 0 \models d'_1 = c +  c_1 \cdot d_1 + \ldots + c_8 \cdot d_8$
for all $1 \leq i \leq \ell$.
The result follows since $\mathbf{M}$ is the least upper bound of $\mathbf{M}_1$,  $\mathbf{M}_2$, \ldots, $\mathbf{M}_\ell$
whereas $d'_1 = c +  c_1 \cdot d_1 + \ldots + c_8 \cdot d_8$ is an upper bound.
\qed


\section{Deriving Updates with Templates}
\label{section:template_octahedra}

The previous section showed how linear equalities can be used to
relate a symbolic constant of an inequality in the output octagon to the
symbolic constants on the inequalities of the input octagon.  In this
section we develop complementary techniques for updates that
cannot be characterised in this way.  To illustrate the problem, Sect.~\ref{section:bounds}
introduces an example which demonstrates why it can be propitious to base updates on
symbolic bounds (range) constraints. Then, Sect.~\ref{section:octagons} refines this observation, demonstrating
the role of octagonal inequalities in constructing updates, while
Sect.~\ref{section:templates} shows how equality constraints can be
combined with auxiliary variables \cite{BRZ05,Col04,MS04}, to derive non-linear relationships
between an output constant
and the symbolic input constants.  These techniques all share the use of templates, either in
the syntactic form of the linear inequalities, or the terms that arise in the non-linear equalities.

\subsection{Bounds constraints}\label{section:bounds}

To illustrate the problem with affine updates, consider the following code block: 
\[
\begin{array}{r@{\;}l@{\qquad}r@{\;}l@{\qquad}r@{\;}l@{\qquad}r@{\;}l}
1: & \texttt{AND R0 15}; &
2: & \texttt{AND R1 15}; &
3: & \texttt{XOR R0 R1}; &
4: & \texttt{ADD R0 R1};
\end{array}
\]
The operations \texttt{AND} and \texttt{XOR} are uni-modal; \texttt{ADD} is multi-modal but it only operates in
the exact non-negative mode for
this block. Since the \texttt{AND} instructions truncate to contents of \texttt{R0} and \texttt{R1}
to the values stored in their low bytes (an operation which is non-linear),
no affine relationship exists
between the symbolic
constants $d_i$ that characterise the input octagon
and those $d'_i$ that characterise the output octagon.
However, observe that it is still possible to find a bound on $d'_1$. In fact, range refinement, as
detailed in Sect.~\ref{section:deriving-guards-sat}, can be applied
to maximise $\signed{\vec{r0}'}$ to infer $\signed{\vec{r0}'} \leq 30$, hence the update 
$d'_1 = 30$.  Repeating this tactic for remaining
the symbolic output constants yields:
\[
\begin{array}{l@{\qquad}l@{\qquad}l@{\qquad}l}
\begin{array}{rcl}
d'_1 & = & 30 \\ 
d'_2 & = & 15 \\ 
\end{array}
&
\begin{array}{rcl}
d'_3 & = & 0 \\ 
d'_4 & = & 0 
\end{array}
&
\begin{array}{rcl}
d'_5 & = & 45 \\ 
d'_6 & = & 0 \\ 
\end{array}
&
\begin{array}{rcl}
d'_7 & = & 0 \\ 
d'_8 & = & 15 
\end{array}
\end{array}
\]

\subsection{Octagonal inequality constraints}\label{section:octagons}

Ranges are merely a degenerate form of octagonal inequality, which suggests using octagons
to relate an output $d'_i$ to an input $d_j$.  To illustrate this idea,
consider the following code that 
rounds \texttt{R0} up the next multiple of 16:
\[
\begin{array}{r@{\;}l@{\qquad}r@{\;}l@{\qquad}r@{\;}l@{\qquad}r@{\;}l}
1: & \texttt{MOV R1 R0}; &
2: & \texttt{NEG R1}; &
3: & \texttt{AND R0 15}; &
4: & \texttt{ADD R0 R1}; \\
\end{array}
\]

\noindent
The instruction \texttt{NEG R1} computes the two's complement of \texttt{R1}, updating
\texttt{R1} with the result.   The instructions \texttt{NEG R1} and \texttt{ADD R0, R1} are multi-modal, thus
consider the feasible mode in which both instructions are exact, the former
being negative and the latter non-negative.
To search for a relationship between $d'_1$ and $d_1$,
the expression  $\signed{\vec{r0}'} - \signed{\vec{r0}}$ is maximised to infer
$\signed{\vec{r0}'} - \signed{\vec{r0}}  \leq 15$, hence
$\signed{\vec{r0}'} \leq \signed{\vec{r0}}  + 15$ thus the update $d'_1 = d_1 + 15$.
Maximising the remaining expressions
\[
\begin{array}{l}
\\
\signed{\vec{r0}'} - (+\signed{\vec{r1}}) \\
\signed{\vec{r0}'} - (-\signed{\vec{r0}}) \\
\signed{\vec{r0}'} - (-\signed{\vec{r1}}) 
\end{array}
\qquad
\begin{array}{l}
\signed{\vec{r0}'} - (+\signed{\vec{r1}} + \signed{\vec{r0}}) \\
\signed{\vec{r0}'} - (-\signed{\vec{r1}} - \signed{\vec{r0}}) \\
\signed{\vec{r0}'} - (-\signed{\vec{r1}} + \signed{\vec{r0}}) \\
\signed{\vec{r0}'} - (+\signed{\vec{r1}} - \signed{\vec{r0}}) 
\end{array}
\]
in general, derives invariants of the form $d'_1 \leq d_2 + c$, \ldots, $d'_1 \leq d_7 + c$
where $c$ is some constant, either of which can be strengthened to an equality and interpreted as an update.  However, in this case, these additional updates
do not yield any further useful bounds on $d'_1$.  Observe too that some of the above
expressions involve 3 variables, whereas some of expressions that bound $d'_5, d'_6, d'_7$ and $d'_8$
involve 4 variables, even though the updates are themselves octagonal.  Completing this derivation
for the above example yields:
\[
\begin{array}{l@{\qquad\qquad}l}
\begin{array}{rcl}
d'_1 & = & d_1 + 15 \\ 
d'_2 & = & 15 \\ 
d'_3 & = & d_3 \\ 
d'_4 & = & 0 
\end{array}
&
\begin{array}{rcl}
d'_5 & = & d_1 + 30 \\ 
d'_6 & = & d_3 \\ 
d'_7 & = & d_3 + 15 \\ 
d'_8 & = & d_1 + 15
\end{array}
\end{array}
\]

\subsection{Non-linear equality constraints}\label{section:templates}
Relaxing the equalities to inequalities provides one degree of freedom for generalising updates; 
relaxing linear equalities to non-linear ones provides another.
Polynomial extensions~\cite[Sect.~6]{MS04} have been proposed for
generalising linear equality analysis, and there is no reason why
this technique cannot be adapted to the problem of deriving transfer functions.

\subsubsection{Generating non-linear equality constraints}

The idea is to augment the original variables in the block with fresh variables
specifically introduced to denote non-linear terms.  The terms are
drawn from a finite language of templates that typically includes monomials up to a fixed degree.
To illustrate, consider the following basic block which computes the location an offset
relative to the start location of two-dimension array where the registers \texttt{R0} and \texttt{R1}
represent the row and column coordinates (which are indexed from 0).  
Register \texttt{R2} represents row size; all registers are signed.
\[
\begin{array}{r@{\;}l@{\qquad}r@{\;}l@{\qquad}r@{\;}l@{\qquad}r@{\;}l}
1: & \texttt{MUL R0 R2}; &
2: & \texttt{ADD R0 R1};
\\
\end{array}
\]
Assume
the block is described as a Boolean formula $\varphi(\vec{X})$ and all operations are exact.
As before, the values of \texttt{R0}, \texttt{R1} and \texttt{R2} on input are represented using
bit-vectors $\vec{r0}$, $\vec{r1}$ and $\vec{r2}$, whereas $\vec{r0}'$ denotes the value of
\texttt{R0} on output. Passing $\varphi(\vec{X})$ to a solver yields a model $\mathbf{m}_1$
as before, say:
\[
\mathbf{m}_1 = 
\left\{
\begin{array}{llllllll}
\signed{\vec{r0}} = 2, & \hspace{1.0em} & 
\signed{\vec{r1}} = 4, & \hspace{1.0em} &
\signed{\vec{r2}} = 3, & \hspace{1.0em} &
\signed{\vec{r0}'} = 10
\end{array}
\right\}
\]
Instead of directly representing these values as an affine system, auxiliary variables are introduced whose
sole purpose is to represent some non-linear terms drawn from a set of templates. In this case,
we have a set that contains the single non-linear term $\signed{\vec{r0}} \cdot \signed{\vec{r2}}$, hence
we introduce a fresh variable $s$ defined
as $s = \signed{\vec{r0}} \cdot \signed{\vec{r2}}$. Since $\signed{\vec{r0}} = 2$ and $\signed{\vec{r2}} = 3$, it follows $s = 6$. 
With the variable ordering $\langle \vec{r0}', \vec{r0}, \vec{r1}, \vec{r2}, s \rangle$ on columns, we obtain the following
affine system $\mathbf{M}_1 \in \mathbb{Z}^{5 \times 6}$ as follows:
\[
\begin{array}{lll}
\mathbf{M}_1 & = & \left[
\begin{array}{ccccc|l}
1 & 0 & 0 & 0 & 0 & 10 \\
0 & 1 & 0 & 0 & 0 & 2 \\
0 & 0 & 1 & 0 & 0 & 4 \\
0 & 0 & 0 & 1 & 0 & 3 \\
0 & 0 & 0 & 0 & 1 & 6
\end{array}
\right]
\end{array}
\]
Now the procedure proceeds much like before.
The formula $\varphi(\vec{X})$ is now augmented with the constraint $\signed{\vec{r0}} \cdot \signed{\vec{r2}} \neq 6$, the resulting
formula being denoted $\varphi'(\vec{X})$. (Propositional encodings have been suggested for 
systems of inequality constraints over 
polynomial terms whose size is quadratic in the
number of symbols required to define the constraints~\cite[Theorem~7]{FGM+07}.)
Passing $\varphi'(\vec{X})$ to a SAT solver yields a model $\mathbf{m}_2$:
\[
\mathbf{m}_2 = 
\left\{
\begin{array}{llllllll}
\signed{\vec{r0}} = 3, & \hspace{1.0em} & 
\signed{\vec{r1}} = 4, & \hspace{1.0em} &
\signed{\vec{r2}} = 8, & \hspace{1.0em} &
\signed{\vec{r0}'} = 28
\end{array}
\right\}
\]
which implies that $s = 24$. The merge is then computed thus:
\[
\begin{array}{lll}
\mathbf{M}_1 \sqcup \mathbf{M}_2 & = & \left[
\begin{array}{ccccc|l}
1 & 0 & 0 & 0 & 0 & 10 \\
0 & 1 & 0 & 0 & 0 & 2 \\
0 & 0 & 1 & 0 & 0 & 4 \\
0 & 0 & 0 & 1 & 0 & 3 \\
0 & 0 & 0 & 0 & 1 & 6
\end{array}
\right] 
\sqcup
\left[
\begin{array}{ccccc|l}
1 & 0 & 0 & 0 & 0 & 28 \\
0 & 1 & 0 & 0 & 0 & 3 \\
0 & 0 & 1 & 0 & 0 & 4 \\
0 & 0 & 0 & 1 & 0 & 8 \\
0 & 0 & 0 & 0 & 1 & 24
\end{array}
\right] \\
\\[-2.5ex]
& = &
\left[
\begin{array}{ccccc|l}
1 & -18 & 0 & 0 & 0 & -26 \\
0 & 5 & 0 & -1 & 0 & 7 \\
0 & 0 & 1 & 0 & 0 & 4 \\
0 & 0 & 0 & 18 & -5 & 24
\end{array}
\right]
\end{array}
\]
and the formula further augmented thus:
\[
\varphi''(\vec{X}) = \varphi'(\vec{X}) \wedge (18 \cdot \signed{\vec{r2}} - 5 \cdot \signed{\vec{r0}} \cdot \signed{\vec{r2}} \neq 24)
\]
This formula gives another model:
\[
\mathbf{m}_3 = 
\left\{
\begin{array}{lllllll}
\signed{\vec{r0}} = 2, & \hspace{1.0em} & 
\signed{\vec{r1}} = 2, & \hspace{1.0em} &
\signed{\vec{r2}} = 2, & \hspace{1.0em} &
\signed{\vec{r0}'} = 6
\end{array}
\right\}
\]
From this we deduce $s = 4$ and hence obtain:
\[
\begin{array}{lll}
\mathbf{M}_1 \sqcup \mathbf{M}_2 \sqcup \mathbf{M}_3 & = & \left[
\begin{array}{ccccc|l}
1 & -18 & 0 & 0 & 0 & -26 \\
0 & 5 & 0 & -1 & 0 & 7 \\
0 & 0 & 1 & 0 & 0 & 4 \\
0 & 0 & 0 & 18 & -5 & 24
\end{array}
\right]
\sqcup
\left[
\begin{array}{ccccc|l}
1 & 0 & 0 & 0 & 0 & 6 \\
0 & 1 & 0 & 0 & 0 & 2 \\
0 & 0 & 1 & 0 & 0 & 2 \\
0 & 0 & 0 & 1 & 0 & 2 \\
0 & 0 & 0 & 0 & 1 & 4
\end{array}
\right] \\
\\[-3ex]
& = &
\left[
\begin{array}{ccccc|l}
1 & -18 & -2 & 0 & 0 & -34 \\
0 & 10 & 1 & -2 & 0 & 18 \\
0 & 0 & 4 & -18 & 5 & -8
\end{array}
\right]
\end{array}
\]
By passing:
\[
\varphi''(\vec{X}) \wedge (4 \cdot \signed{\vec{r1}} - 18 \cdot \signed{\vec{r2}} + 5 \cdot\signed{\vec{r0}} \cdot \signed{\vec{r2}}) \neq -8
\]
to the solver, we generate yet another model:
\[
\mathbf{m}_3 = 
\left\{
\begin{array}{llllllll}
\signed{\vec{r0}} = 4, & \hspace{1.0em} & 
\signed{\vec{r1}} = 5, & \hspace{1.0em} &
\signed{\vec{r2}} = 2, & \hspace{1.0em} &
\signed{\vec{r0}'} = 13
\end{array}
\right\}
\]
Joining $\mathbf{M}_4$ with $\mathbf{M}_1 \sqcup \mathbf{M}_2 \sqcup \mathbf{M}_3$ then produces the system:
\[
\begin{array}{lll}
\mathbf{M}_1 \sqcup \mathbf{M}_2 \sqcup \mathbf{M}_3 \sqcup \mathbf{M}_4 & = &
\left[
\begin{array}{ccccc|l}
12 & -64 & 0 & -70 & 10 & -152 \\
0 & 64 & -12 & 70 & -23 & 152
\end{array}
\right]
\end{array}
\]
Proceeding with one more iteration, we obtain the system:
\[
\begin{array}{lll}
\mathbf{M}_1 \sqcup \mathbf{M}_2 \sqcup \mathbf{M}_3 \sqcup \mathbf{M}_4 \sqcup \mathbf{M}_5 & = &
\left[
\begin{array}{ccccc|l}
1 & 0 & -1 & 0 & -1 & 0
\end{array}
\right]
\end{array}
\]
Since adding another disequality constraint yields an unsatisfiable system, the equation:
\[
\signed{\vec{r0}'} = \signed{\vec{r1}} + s = \signed{\vec{r1}} + \signed{\vec{r0}} \cdot \signed{\vec{r2}}
\]
characterises the polynomial input-output relation implemented by this block. 
Observe that the total number of calls to a SAT solver is still linear in the number of
overall variables --- program variables plus auxiliary variables --- due to linear chain lengths in affine
spaces. Our prototype implementation written in \textsc{Java} on top of the \textsc{[mc]square} framework~\cite{Sch09}
and \textsc{Sat4J}~\cite{sat4j} computes this update in no more than $0.25$s.

\subsubsection{Lifting non-linear equality constraints to intervals}

Of course, as in the Sect.~\ref{sect-linear-affine}, the derived constraint relates neither internal bounds nor
symbolic constants on the input and output octagons.
One would expect that non-linear equality constraints can be straightforwardly
lifted to updates using either the technique
described in Sect.~\ref{section-medium} or by using the
maximisation technique explained in Sect.~\ref{section-exact}. However, this is not so.

To illustrate, consider polynomial extension
applied to interval relations, and in particular the problem
of lifting the above affine system to construct an update over
the symbolic
bounds $ \vec{r0}_{\mymin}$, $\vec{r0}_{\mymax}$,  $\vec{r1}_{\mymin}$, $\vec{r1}_{\mymax}$, 
$\vec{r2}_{\mymin}$, $\vec{r2}_{\mymax}$, $\vec{r0}'_{\mymin}$ and $\vec{r0}'_{\mymax}$.
Observe that the original equation
\[
\signed{\vec{r0}'} = \signed{\vec{r1}} + \signed{\vec{r0}} \cdot \signed{\vec{r2}}
\]
gives rise to two updates:
\[
\begin{array}{@{}l@{\;}l@{\;}l@{}}
\signed{\vec{r0}'_{\mymin}} & = & \signed{\vec{r1}_{\mymin}} + \min \{
\signed{\vec{r0}_{\mymin}} \cdot \signed{\vec{r2}_{\mymin}},
\signed{\vec{r0}_{\mymin}} \cdot \signed{\vec{r2}_{\mymax}},
\signed{\vec{r0}_{\mymax}} \cdot \signed{\vec{r2}_{\mymin}},
\signed{\vec{r0}_{\mymax}} \cdot \signed{\vec{r2}_{\mymax}} \} \\
\signed{\vec{r0}'_{\mymax}} & = & \signed{\vec{r1}_{\mymax}} + \max \{
\signed{\vec{r0}_{\mymin}} \cdot \signed{\vec{r2}_{\mymin}},
\signed{\vec{r0}_{\mymin}} \cdot \signed{\vec{r2}_{\mymax}},
\signed{\vec{r0}_{\mymax}} \cdot \signed{\vec{r2}_{\mymin}},
\signed{\vec{r0}_{\mymax}} \cdot \signed{\vec{r2}_{\mymax}} \} 
\end{array}
\]
that involve, respectively, minimisation and maximisation operations.  These operations
are required because
it is not until the symbolic bounds are instantiated
that the relative sizes of the non-linear terms can be compared. 
(These comparisons are redundant for linear terms because they are monotonic.)

To present this transformation formally, let $\vec{V} \cup \vec{V}'$ denote the
input and output variables, and $S$ denote a set of templates (monomials) over the variables $\vec{V}$.
Thus if $s \in S$ then $s = \Pi_{i=1}^{n} v_i$ for some $v_i \in \vec{V}$.
We introduce a map $\mu(s) = \{  \Pi_{i=1}^{n} w_i \mid w_i = {v_i}_{\mymin} \vee w_i = {v_i}_{\mymax} \}$
so that, for example, if $s = \signed{\vec{r0}} \cdot \signed{\vec{r2}}$ then:
\[
\mu(\signed{\vec{r0}} \cdot \signed{\vec{r2}}) =  \{ 
	\signed{\vec{r0}_{\mymin}} \cdot \signed{\vec{r2}_{\mymin}},
	\signed{\vec{r0}_{\mymin}} \cdot \signed{\vec{r2}_{\mymax}},
	\signed{\vec{r0}_{\mymax}} \cdot \signed{\vec{r2}_{\mymin}},
	\signed{\vec{r0}_{\mymax}} \cdot \signed{\vec{r2}_{\mymax}} 
	\}
\]

\noindent Each of the polynomially extended equations take the form:
\[
\begin{array}{lll}
\lambda_{\vec{v}'} \cdot \vec{v}' & = & \sum_{\vec{v} \in \vec{V}} \lambda_{\vec{v}} \cdot \vec{v} + \sum_{s \in \mathcal{S}} \lambda_{s} \cdot s + d
\end{array}
\]
where $\vec{v}' \inÊ\vec{V}'$, $\lambda_{\vec{v}'} \in \mathbb{N}$, and
$\lambda_{\vec{v}} \in \mathbb{Z}$ for all $\vec{v} \inÊ\vec{V}$, and
$\lambda_{s} \inÊ\mathbb{Z}$ for all $s \in S$. 

We then replace each polynomially extended
equation by a pair of equations as follows:
\[
\begin{array}{lllllllll}
\lambda_{\vec{v}'} \cdot \vec{v}'_{\mymin} & = & \sum_{\vec{v} \inÊ\vec{V}} \lambda_{\vec{v}} \cdot \beta(-\lambda_{\vec{v}}, \vec{v}) & + & \sum_{s \inÊ\mathcal{S}} \lambda_{s} \cdot \gamma(-\lambda_s, s) & + & d \\
\lambda_{\vec{v}'} \cdot \vec{v}'_{\mymax} & = & \sum_{\vec{v} \inÊ\vec{V}} \lambda_{\vec{v}} \cdot \beta(\lambda_{\vec{v}}, \vec{v}) & + & \sum_{s \inÊ\mathcal{S}} \lambda_{s} \cdot \gamma(\lambda_s, s) & + & d
\end{array}
\]
where $\beta$ is defined as before (in Sect.~\ref{section:worked_examples:lifting}) and
$\gamma$ transforms the monomials as follows:
\[
\begin{array}{lll}
\gamma(\lambda, s) & = &
	\left\{
		\begin{array}{ll}
			\min(\mu(s)) & : \text{ if } \lambda < 0 \\
			\max(\mu(s)) & : \text{ otherwise}
		\end{array}
	\right.
\end{array}
\]
Note that linear terms are transformed in the same manner as before.

\subsubsection{Non-linear equality constraints and octagons}

The minimisation and maximisation terms that arise in interval updates suggest
a tactic for inferring updates for octagons in the presence of non-linear terms.
To illustrate with the above example, the construction
proceeds by
introducing fresh variables $s_1$ and $s_2$ defined such that:
\[
s_1 = \max(d_1 \cdot d_3, d_1 \cdot d_6, d_5 \cdot d_3, d_5 \cdot d_6 ) \qquad
s_2 = \min(d_1 \cdot d_3, d_1 \cdot d_6, d_5 \cdot d_3, d_5 \cdot d_6 )
\]
Then maximisation is interleaved with affine join, as detailed in Sect.~\ref{section-exact}, so
as to derive updates between a $d'_i$ variable, the $d_j$ and the auxiliary
$s_1$ and $s_2$ variables.
By applying this technique the following transfer function is derived:
\[
\begin{array}{lll}
\left\{
\begin{array}{rllllll}
\signed{\vec{r0}} & \leq & d_1 \\
\signed{\vec{r1}} & \leq & d_2 \\
\signed{\vec{r2}} & \leq & d_3 \\
-\signed{\vec{r0}} & \leq & d_4 \\
-\signed{\vec{r1}} & \leq & d_5 \\
-\signed{\vec{r2}} & \leq & d_6 \\
\hline
\signed{\vec{r0}} + \signed{\vec{r1}} & \leq & d_7 \\
-\signed{\vec{r0}} - \signed{\vec{r1}} & \leq & d_8 \\
-\signed{\vec{r0}} + \signed{\vec{r1}} & \leq & d_9 \\
\signed{\vec{r0}} - \signed{\vec{r1}} & \leq & d_{10} \\
\hline
\signed{\vec{r0}} + \signed{\vec{r2}} & \leq & d_{11} \\
-\signed{\vec{r0}} - \signed{\vec{r2}} & \leq & d_{12} \\
-\signed{\vec{r0}} + \signed{\vec{r2}} & \leq & d_{13} \\
\signed{\vec{r0}} - \signed{\vec{r2}} & \leq & d_{14} \\
\hline
\signed{\vec{r1}} + \signed{\vec{r2}} & \leq & d_{15} \\
-\signed{\vec{r1}} - \signed{\vec{r2}} & \leq & d_{16} \\
-\signed{\vec{r1}} + \signed{\vec{r2}} & \leq & d_{17} \\
\signed{\vec{r1}} - \signed{\vec{r2}} & \leq & d_{18} \\
\end{array}
\right\}
&
\hspace{1em}
\leadsto
\hspace{1em}
&
\left\{
\begin{array}{rllllll}
\signed{\vec{r0}'} & \leq & d_2  + s_1  \\
\signed{\vec{r1}'} & \leq & d_2 \\
\signed{\vec{r2}'} & \leq & d_3 \\
-\signed{\vec{r0}'} & \leq & d_5 + s_2 \\
-\signed{\vec{r1}'} & \leq & d_5 \\
-\signed{\vec{r2}'} & \leq & d_6 \\
\hline
\signed{\vec{r0}'} + \signed{\vec{r1}'} & \leq & d_2 + s_1 + d_2 \\
-\signed{\vec{r0}'} - \signed{\vec{r1}'} & \leq & d_5 + s_2 + d_5 \\
-\signed{\vec{r0}'} + \signed{\vec{r1}'} & \leq & d_5 + s_2 + d_2 \\
\signed{\vec{r0}'} - \signed{\vec{r1}'} & \leq & d_2  + s_1 + d_5 \\
\hline
\signed{\vec{r0}'} + \signed{\vec{r2}'} & \leq & d_2  + s_1 + d_3 \\
-\signed{\vec{r0}'} - \signed{\vec{r2}'} & \leq & d_5 + s_2 + d_6 \\
-\signed{\vec{r0}'} + \signed{\vec{r2}'} & \leq & d_5 + s_2 + d_3 \\
\signed{\vec{r0}'} - \signed{\vec{r2}'} & \leq & d_2  + s_1 + d_6 \\
\hline
\signed{\vec{r1}'} + \signed{\vec{r2}'} & \leq & d_{15} \\
-\signed{\vec{r1}'} - \signed{\vec{r2}'} & \leq & d_{16} \\
-\signed{\vec{r1}'} + \signed{\vec{r2}'} & \leq & d_{17} \\
\signed{\vec{r1}'} - \signed{\vec{r2}'} & \leq & d_{18} \\
\end{array}
\right\}
\end{array}
\]
Thus, for example, if the octagonal describes a cube that is offset
from the origin, namely
$d_1 = d_2 = d_3 = 3$ and $d_4 = d_5 = d_6 = -2$,
then bound on
$\signed{\vec{r0}'} $, denoted $d'_1$, is calculated by
$d'_1 = d_2  + s_1$ = $3 + \max(3 \cdot 3, 3 \cdot -2, -2 \cdot 3, -2 \cdot -2) = 12$. 


\section{Evaluating Transfer Functions}\label{section:evaluating}

Thus far, we have described how to derive transfer functions for intervals
and octagons where the functions are systems of guards
paired with affine updates, without reference
to how they are evaluated. In our previous
work~\cite{BK10}, the application of
a transfer function amounted to solving a series of
integer linear programs (ILPs).
To illustrate, suppose a transfer function
consists of a single guard $g$ and update $u$ pair
and let
$c$ denote a system of octagonal constraints on the input variables.
A single output inequality in the output system, $c'$, such as
$r0' + r1' \leq d'_5$, can be derived
by maximising $r0' + r1'$ subject to the linear system
$c \wedge g \wedge u$.  To construct $c'$ in its entirety requires
the solution of $O(n^2)$ ILPs where $n$ is the number of
registers (or variables) in the block. Although steady progress has been made
on deriving safe bounds for integer programs \cite{neumaier04safe}, a
more attractive solution computationally would avoid ILPs altogether.

\subsection{A single guard and update pair}
\label{section:single-guard-and-update}

Affine updates, as derived in Sect.~\ref{section-exact},
relate symbolic
constants on the inequalities in the input octagon to those of the output octagon.
These updates confer a different, simpler, evaluation model.
To compute $r0' + r1' \leq d'_5$ in $c'$ it is sufficient to
compute $c \sqcap g$ \cite{Min06} which is the octagon
that describes the conjoined system $c \wedge g$.  This
can be computed in quadratic-time when $g$
is a single inequality and in cubic-time
otherwise \cite{Min06}. The meet
$c \sqcap g$ then defines values for the symbolic
constants $d_i$, though these values may include $-\infty$ and $\infty$.  The value of
$d'_5$ is defined by its affine update, that is, as a weighted
sum of the $d_i$ values.  If there is no affine update for $d'_5$, then its value
defaults to $\infty$.  If bounds have been inferred for output octagons,
then the $d'_i$ can possibly be refined with a tighter bound. 
This evaluation mechanism thus
replaces ILP with arithmetic that is both conceptually
simple and computationally efficient.  This is significant
since transfer functions are themselves computed
many times during fixed point evaluation.

\subsection{A system of guard and update pairs}

The above evaluation procedure needs to be applied for each
guard $g$ and update $u$ pair
for which $c \sqcap g$ is satisfiable.  Thus
several output octagons may be derived for a single block.
We do not prescribe how these octagons should be combined,
for example,
a disjunctive representation is one possibility~\cite{GR98}.
However, the simplest tactic is undoubtedly to apply the
merge operation for octagons \cite{Min06} (though this entails
closing the output octagons).

\subsection{A system of updates for template inequality constraints}

Evaluating an octagonal update represented as an affine
equality as discussed in Sect.~\ref{section:single-guard-and-update}
is straightforward since each symbolic bound $d'_i$ on output is
characterised by exactly one linear equation. This is not necessarily
the case if template inequality constraints have been applied to derive updates, as
discussed in Sect.~\ref{section:octagons}. Recall, for example,
that inequalities of the form
$d'_1 \leq d_2 + c$, \ldots, $d'_1 \leq d_7 + c$ can arise, all of
which potentially induce non-trivial bounds on $d'_1$.  In general,
a symbolic constant $d'_i$ in the output octagon might be
related to the input symbolic constants $d_1, \ldots, d_n$ through
a system of $m$ inequalities:
\[
\bigwedge_{j=1}^m \left( c_j d'_i \leq \sum_{k=1}^n c_{j,k} \cdot d_k \right)
\]
where $c_j > 0$ otherwise the inequality does not bound $d'_i$ from above and can
thus be discarded. Although any of these inequality can be strengthened to an equality and
interpreted as an update, it is more
precise to compute:
\[
d'_i = \min \left\{ \left. \left\lfloor \left(\sum_{k=1}^n c_{j,k} \cdot d_k \right) /c_j \right\rfloor  \right| 1 \leq j \leq m \right\}
\]
Therefore, in general, transfer function evaluation can involve the evaluation of several linear
expressions for each symbolic constant in the output octagon. 


\section{Experiments}
\label{section:experiments}


We have implemented the techniques described in this paper in
\textsc{Java} using the \textsc{Sat4J} solver~\cite{sat4j}, so as to
integrate with our analysis framework for machine code~\cite{Sch09}, called
\textsc{[mc]square}, which is also coded in \textsc{Java}. All
experiments were performed on a \textsc{MacBook Pro} equipped with a
2.6 GHz dual-core processor and 4 GB of RAM, but only a single core
was used in our experiments.

To evaluate transfer function synthesis without quantifier elimination,
Tab.~\ref{table:experimental_results} compares the results for intervals
for different blocks of assembly code to those obtained using the technique
described in~\cite{BK10}. This corresponds to the techniques
presented in Sect.~\ref{section:guards} and Sect.~\ref{section:updates}.
Column \emph{\#instr} contains the number of
instructions, whereas column \emph{\#bits}
gives the bit-width. (The 
8-bit and 32-bit versions of the AVR instruction sets are
analogous.)
Then, \emph{\#affine}
presents the number of affine relations for each block. The columns
\emph{runtime} contain the runtime and the number of SAT instances. 
The overall runtime of the elimination-based algorithm~\cite{BK10}
is given in column \emph{old} ($\infty$ is used for timeout,
which is set to 30s). Transfer function synthesis for blocks of up to 10
instruction is evaluated, which is a typical size for microcontroller code.
For these size blocks, we have never observed more than 10
feasible mode combinations.

\begin{table}[!t]
\caption{Experimental results for synthesis of transfer functions}
\label{table:experimental_results}
\centering
\begin{tabular}{|l||c|c|c||c|c|c||c|}
\hline
\multirow{2}{*}{block} & \multirow{2}{*}{\#instr} & \multirow{2}{*}{\#affine} & \multirow{2}{*}{\#bits} & \multicolumn{4}{|c|}{runtime} \\ \cline{5-8}
 & & & & guards / \#SAT & affine / \#SAT & overall & old \\
\hline
\hline
\multirow{2}{*}{\texttt{inc}} & \multirow{2}{*}{1} & \multirow{2}{*}{2}  & 8 & 0.2s / 32 & 0.1s / 5 & 0.3s & 0.2s \\
& & & 32 & 0.5s / 128 & 0.2s / 5 & 1.0s & 23.0s \\
\hline
\multirow{2}{*}{\texttt{inc+shift}} & \multirow{2}{*}{2} & \multirow{2}{*}{3}  & 8 & 0.3s / 48 & 0.1s / 8 & 0.4s & 0.3s \\
& & & 32 & 0.8s / 192 & 0.2s / 8 & 1.0s & $\infty$ \\
\hline
\multirow{2}{*}{\texttt{swap}} & \multirow{2}{*}{3} & \multirow{2}{*}{1} & 8 & --- & 0.1s / 3 & 0.1s & 0.1s \\
& & & 32 & --- & 0.1s / 3 & 0.1s & 0.2s \\
\hline
\multirow{2}{*}{\texttt{inc+flip}} & \multirow{2}{*}{4} & \multirow{2}{*}{2} & 8 & 0.2s  / 32 & 0.2s / 5 & 0.4s & 0.5s \\
& & & 32 & 0.9s / 128 & 0.3s / 5 & 1.2s & $\infty$ \\
\hline
\multirow{2}{*}{\texttt{abs}} & \multirow{2}{*}{5} & \multirow{2}{*}{3} & 8 & 2.5s / 216 & 0.3s / 8 & 2.8s & 0.8s \\
& & & 32 & 6.5s / 792 & 0.3s / 8 & 6.8s & $\infty$ \\
\hline
\multirow{2}{*}{\texttt{inc+abs}} & \multirow{2}{*}{6} & \multirow{2}{*}{3} & 8 & 2.6s / 216 & 0.3s / 8 & 2.9s & 1.4s \\
& & & 32 & 6.7s / 792 & 0.3s / 8 & 7.0s & $\infty$ \\
\hline
\multirow{2}{*}{\texttt{sum+isign}} & \multirow{2}{*}{7} & \multirow{2}{*}{9} & 8 & 5.9s / 648 & 0.2s / 24 & 4.3s & 4.5s \\
& & & 32 & 19.7s / 2376 & 0.4s / 24 & 11.1s & $\infty$ \\
\hline
\texttt{exchange+} & \multirow{2}{*}{10} & \multirow{2}{*}{3} & 8 & 2.8s / 216 & 0.3s / 8 & 3.1s & 9.5s \\
\texttt{abs} & & & 32 & 7.2s / 792 & 0.3s / 8 & 7.5s & $\infty$ \\
\hline
\end{tabular}
\end{table}

\subsection{Comparison}

Using quantifier elimination, all instances could be solved in a
reasonable amount of time for 8-bit instructions. However, only the small instances could be solved for 32 bits
(and only then because the Boolean encodings for the instructions
were minimised prior to the synthesis of the transfer functions).
It is also important to appreciate that none of the timeouts was caused by the SAT solver;
it was resolution that failed to produce results in reasonable time.
By way of comparison, synthesising guards for different overflow modes
requires most runtime in our new
approach, caused by the fact that the number of SAT instances to be solved
grows linearly with the number of bits and quadratically with
the number of variables
(the number of octagonal inequalities is quadratic in the number of variables).
Computing the affine updates consumes only a fraction of the overall time. In
terms of precision, the results coincide with those previously generated \cite{BK10}.

The block for \texttt{swap} is interesting since it
consists of three consecutive exclusive-or instructions, for which there is no
coupling between different bits of the same register. 
The block is also unusual in that it is uni-modal with vacuous guards.
These properties make it ideal for resolution.  Even in this situation, the new
technique scales better. In fact,  
the Boolean formulae that we present to the solver 
are almost trivial by modern standards, the main overhead coming from
repeated SAT solving rather than solving a single large instance.
{\sc Sat4J} does reuse clauses learnt in an earlier SAT instances,
though it does not permit clauses to be incrementally added and rescinded 
which is useful when solving maximisation problems~\cite{BK10}.
Thus the timings given above are very conservative; indeed
{\sc Sat4J} was chosen to maintain
the portability of {\sc [mc]square} rather than for raw performance.
Nevertheless, these timings very favourably compare with those required
to compute transfer functions for intervals using BDDs~\cite{regehr04hoist}, where in excess of 24 hours is required for
single 8-bit instructions. Our experiences~\cite{BK10b,RB11} with native solvers such as
\textsc{MiniSat}, however, indicate that a tenfold speed-up can be achieved
by replacing \textsc{Sat4J}.

\subsection{Deriving octagonal transfer functions} The process of
deriving octagonal transfer functions by lifting
(Sect.~\ref{section-medium}) requires
an imperceivable overhead compared to computing affine relations themselves,
indeed it is merely syntactic rewriting.
The runtimes required for inferring
affine updates
by alternating range refinement and affine join
(Sect.~\ref{section-exact}), however, is typically 3 or 4 times
slower than those of computing the guards;
the number of symbolic constants on the output inequalities
corresponds exactly to the number of input guards.
Since the octagon on input consists of 8 guards, and so does
the octagon on output, the worst case requires $16+1$ iterations
of affine abstraction and refinement; a single iteration of refinement
is no more expensive as in the cases given in Tab.~\ref{table:experimental_results},
and the affine join has imperceivable impact. We have observed the full number
of iterations is only needed for programs for which there is no affine
relation between octagons on input and output. We refrain from giving exact times
for the affine updates since they were computed with {\sc Z3} \cite{z3} rather than {\sc Sat4J} and
thus are not directly comparable.

\subsection{Further optimisations} Since transfer functions
are program dependent, one could first use a simple form of range
analysis~\cite{BK10b,CLS08,RB11} to over-approximate the
ranges a register can assume. These ranges can be
encoded in the formulae, thereby pruning out some mode-combinations.
For example, it is rarely the case that the absolute value function
is actually applied to the smallest representable integer. 


\section{Related Work}
\label{section:related_work}


The problem of designing transfer functions for numeric domains is as old
as the field of abstract interpretation itself~\cite{CC77}, and even
the technique of using primed and unprimed variables to capture and abstract
the semantics of instructions and functions dates back to the
thesis work of Halbwachs \cite{halbwachs79determination}.
However, even for a fixed abstract domain, there are typically many ways of designing 
and implementing transfer functions.
Cousot and Halbwachs~\cite[Sect.~4.2.1]{CH78}, for example, discussed
several ways to realise a transfer function for assignments such as
$x = y \times z$ in the polyhedral domain while
abstracting integer division
$x = y / z$ is an interesting study within itself \cite{simon08value}.

The problem of handcrafting best transformers is particularly challenging and
Granger~\cite{Gra89} lamented the difficulty of devising precise transfer functions for linear congruences. 
However, it took more than a decade after Granger's work before it was observed that best transformers
could automatically be constructed for domains of finite height~\cite{RSY04}.
Nevertheless, automatic abstraction (or the automatic synthesis of abstractions) has only recently become a
practical proposition, due to emergence of robust decision procedures~\cite{BK10,KS10,Mon09} and efficient quantifier elimination techniques~\cite{BKK11a,KS08b,Mon10}.

\subsection{Generation of symbolic best transformers}

Transfer functions can always be found for domains of finite height using the method of Reps et al.~\cite{RSY04},
provided one is prepared to pay the cost of repeatedly calling a decision procedure or a
theorem prover, possibly many times on each application of a transformer.  
This motivates  applying a decision procedure in order to compute a best transformer offline, prior to the actual
analysis~\cite{BK10,KS10}, so as to both simplify and speedup their application.

Our previous work~\cite{BK10} shows how bit-blasting and quantifier elimination 
can be applied to synthesise transformers for bit-vector programs. This work
was inspired by that of Monniaux~\cite{Mon09,Mon10} on synthesising transfer functions for
piecewise linear programs.  Although his approach extends beyond octagons \cite{simon10tvpi},
it is unclear how to express some instructions (such as bit-wise exclusive-or) in terms of linear constraints.
Universal quantification, as used in both approaches, also appears
in work on inferring linear template constraints~\cite{GSV08}. There, Gulwani and his co-authors
apply Farkas' lemma in order to transform universal quantification into existential quantification, albeit at the
cost of completeness since Farkas' lemma prevents integral reasoning. However,
crucially, neither Monniaux nor Gulwani et al.~provide a way to model integer overflow and underflow. 
Our work explains how to systematically handle wrap-around arithmetic in
the transfer function itself (without having to the revise the notion of abstraction \cite{SK07b})
whilst sidestepping quantifier elimination too.

Transfer functions for low-level code have been synthesised for intervals
using BDDs \cite{bryant86graph} by applying interval subdivision where the extrema
representing the interval are themselves represented as bit-vectors \cite{regehr04hoist}.
If $g : [0, 2^8 -  1] \to [0, 2^8 - 1]$
is a unary operation on an unsigned byte, then its
abstract transformer $f : D \to D$
on $D = \{ \emptyset \} \cup \{ [\ell, u] \mid 0 \leq \ell \leq u < 2^8 \}$
can be defined recursively.
If $\ell = u$ then $f([\ell, u]) = g(\ell)$ whereas
if $\ell < u$ then $f([\ell, u]) = f([\ell, m - 1]) \sqcup f([m,
u])$ where $m = \lfloor u/2^{n} \rfloor 2^n$ and $n = \lfloor
\log_2(u - \ell + 1) \rfloor$.  Binary operations can likewise be
decomposed by repeatedly dividing  squares into their quadrants.  The 8-bit inputs, $\ell$ and $u$, can be represented
as 8-bit vectors, as can the 8-bit outputs, so as to represent $f$
with a BDD.  This permits caching to be applied when $f$ is computed,
which reduces the time needed to compute a best transformer
to approximately 24 hours for each 8-bit operation.   It is difficult to see how
this approach can be extended to blocks that involve many variables
without a step-change in BDD performance.

The question of how to construct a best abstract transformer has also been
considered in the context of Markov decision processes (MDPs) for which
the first abstract interpretation framework has recently been developed \cite{WZ10}. The
framework affords the calculation of both lower and upper bounds on
reachability probabilities, which is novel. The work focuses on predicate abstraction~\cite{GS97b},
that have had some success with large MDPs, and seeks to
answer the question of, for given set of predicates, what is the most precise
abstract program that still is a correct abstraction. More generally, the work illustrates
that the question of how to compute the best abstract transformer is pertinent even in a probabilistic setting.

\subsection{Modular Arithmetic}

The classical approach to handling overflows is to follow the application of a transfer
function with overflow and underflow checks; 
program variables are considered to be
unbounded for the purposes of applying the transfer function but then their sizes are considered
and range tests and, if necessary, range adjustments
are applied to model any wrapping.
This approach has been implemented in the \textsc{Astree} analyzer~\cite{BCC+03,CCF+05}. 
However, for convex polyhedra, it is also possible to revise the concretisation map to reflect
truncation so as to remove the range tests from most abstract operations~\cite{bygde11fully,SK07b}.
Another choice is to deploy congruence relations~\cite{Gra89,Gra97} where the modulus is a power of two
so as to reflect the wrapping in the abstract domain itself \cite{MS07}.  This approach can be applied
to find both relationships between different words \cite{MS07} and the bits that constitute
words  \cite{BKK10,KS08a,KS10} (the relative precision of these two approaches has recently
been compared \cite{elder11abstract}).
Bit-level models have been combined with range inference \cite{BK10b,CLS08}, though neither of these works
address relational abstraction nor transfer function synthesis.

Modular arithmetic can be modelled with case splitting by introducing
a propositional variable that acts as a witness to an overflow. To illustrate,
consider the 8-bit comparison $x + 100 \leq 10$ \cite[Sect.~6.4]{KS08b}.  To
model overflow a witness $p \Leftrightarrow (x + 100 \leq 255)$ is defined, which
is used to control case selection.  Case selection is realised through two constraints
defined by 
$p \Rightarrow (x + 100) \leq 10$
and
$(\neg p) \Rightarrow ((x + 100) - 256) \leq 10$.
Case-based axiomisations can even be used to 
model underflows and rounding-to-zero in IEEE-745 floating-point
arithmetic as shown in~\cite[Sect.~4.5]{Mon09}.
These ideas are similar in spirit to those given in this paper for decomposing
a block into its modes which are selected by guards. 

\subsection{Polynomial Relations}

The last decade has seen increasing interest in the derivation of polynomial invariants, with techniques
broadly falling into two classes:  methods that use algebraic techniques to operate directly over polynomials and
methods that model polynomial invariants in a linear setting. The work of
Col\'{o}n~\cite{Col04} is representative of the latter, for he shows how polynomial relations of bounded degree can be derived
using program transformation. To illustrate, suppose a variable $a$ is updated using the assignment $a = a + 1$. 
A variable $s$ is introduced to represent the non-linear term $a^2$ and the program is extended
by replacing the assignment $a = a+1$  with the parallel assignment
$\langle a, s \rangle = \langle a+ 1, s + 2a + 1 \rangle$ so as
to reflect the update on $a$ to $s$. Linear invariants between $a, s$ and the other variables
in the transformed program then are reinterpreted as polynomial invariants.
The idea of using nonlinear terms as additional independent variables also arises in the work
of Bagnara et al.~\cite{BRZ05} who use convex polyhedra to represent polynomial cones of bounded degree and
thereby derive polynomial inequalities.   They reduce the loss of precision induced through linearisation by
additional linear inequalities, which are included in the polyhedra to express
redundant non-linear constraints.
The idea of extending a vector of variables with non-linear terms also arises in the work
of
M{\"u}ller-Olm and Seidl~\cite{MS04} who consider the complexity of inferring polynomial
equalities up to a fixed degree.  They represent an affine relation with a set of vectors that generate
the space through linear combination. Extending this idea to variables that represent non-linear
terms naturally leads to the notion of polynomial hull which is not dissimilar to the closure
algorithm that is used in this paper for computing non-linear update functions.

Quantifier elimination has been proposed as a technique for inferring polynomial inequalities
directly \cite{Kap05} in which the invariants are templates of
polynomial inequalities with undetermined coefficients. Deriving coefficients for the templates
amounts to applying quantifier elimination
which can be computed using a parametric (or comprehensive) Gr\"{o}bner basis construction~\cite{weispfenning92comprehensive}.
This approach resonates with the technique
proposed by Monniaux for inferring loop invariants~\cite{Mon10b}.
Gr{\"o}bner bases also arise in techniques for calculating invariants that are based on fixed point calculation
\cite{RK04,RK07}, the main advantage of this approach being that it does not assume any a priori bound
on the degree of a polynomial as an invariant. Polynomial analysis has also been applied in the
field of SAT-based termination analysis~\cite{FGM+07} using term rewriting~\cite{GST06,TG03}. Their work provides
techniques for encoding polynomial equality and inequality constraints in propositional Boolean logic.


\subsection{Procedure summaries}

Abstracting the effect of a procedure in a summary is a key problem
in inter-procedural analysis \cite{SP81}  since it enables the effect of
a call on abstract state to be determined without repeatedly tracing the call.
The challenge posed by summaries is how they can be densely represented
whilst supporting the function composition and function application. Gen/kill bit-vector
problems \cite{RHS95precise} are amenable to efficient representation, though
for other problems, such as that of tracking
two variable equalities \cite{MS08}, it is better not to tabular the effect
of a call directly. This is because if a transformer is distributive, then the lower adjoint of a transformer
uniquely determines the transformer and, perhaps surprisingly,
the lower adjoint can sometimes be represented more succinctly than the transformer itself.

Acceleration \cite{GH06,LS07a,LS07b,SJ11} is attracting increasing interest
as an alternative way
of computing a summary of a procedure, or more exactly the loops that it contains.
The idea is to track how program state changes on each loop
iteration so as to compute the trajectory of these changes (in a computation
is that akin to transitive closure) and hence derive, in a single step, a loop invariant that
holds on all iterations of the loop.

Symbolic bounds, which are key to our transfer functions, also arise
in a form of symbolic bounds analysis \cite{RR05} that aspires to infer
ranges on pointer and array index variables
in terms of the parameters of a procedure.  Lower and upper bounds on each
program variable at each program point
are formulated as linear functions of the parameters of the function where the
coefficients are themselves parametric.  The problem then amounts to
inferring values for these parametric coefficients. 
By assuming variables to be non-negative, inequalities between the symbolic bounds can be
reduced to inequalities between the parametric coefficients, thereby reducing the problem
to linear programming.


\section{Concluding Discussion} \label{section:discussion}

\subsection{Synopsis} 

This article discusses the problem of automatically computing transfer functions for programs
whose semantics is defined over finite bit-vectors. The key aspect that distinguishes our work
from existing techniques~\cite{BK10,Mon09,Mon10} is that it does not depend on quantifier
elimination techniques at all. Although Boolean formulae presented in
CNF initially appear attractive for this task because of the simplicity of universal
quantifier elimination~\cite[Sect.~1.3]{BK10}, their real strength is the fact that they are discrete.
This permits linear equalities and inequalities to be inferred by repeated (incremental) satisfiability testing,
avoiding the need for quantifier elimination in the abstraction process entirely.
Most notably, this technique sidesteps the complexity of binary resolution. The force of this observation is
that it extends transfer function synthesis to architectures whose word size exceeds $8$ bits,
thereby strengthening the case for low-level code verification~\cite{BR10,BH08,BH11,BHL+11,BJAS11,FPS11,Sch09,TLL+10}.

\subsection{Future work}

The problem of synthesising transfer functions is not dissimilar to that of inferring ranking functions
for bit-vector relations~\cite{CKRW10}. Given a path $\pi$ with a transition relation $r_{\pi}(\vec{x},\vec{x}')$,
proving the existence of a ranking function amounts to solving the formula
$\exists \vec{c} : \forall \vec{x} : \forall \vec{x}' : r_{\pi}(\vec{x},\vec{x}') \rightarrow (p(\vec{c},\vec{x}) < p(\vec{c},\vec{x}'))$
where $p(\vec{c},\vec{x})$ is a polynomial over the bit-vector $\vec{x}$ and $\vec{c}$ is a bit-vector of coefficients~\cite[Thm.~2]{CKRW10}.
However, if intermediate variables $\vec{y}$ are needed to express
$r_{\pi}(\vec{x},\vec{x}')$, $p(\vec{c},\vec{x})$, $p(\vec{c},\vec{x}')$ or $<$, then the formula actually
takes the form $\exists \vec{c} : \forall \vec{x} : \forall \vec{x}' : \exists \vec{y} : \nu$ where
$\exists \vec{y} : \nu$ is equisatisfiable to $r_{\pi}(\vec{x},\vec{x}') \rightarrow (p(\vec{c},\vec{x}) < p(\vec{c},\vec{x}'))$.
This formula is structurally similar to those solved in~\cite{BK10} by quantifier elimination, which begs the
question of whether this problem --- like that of transfer function synthesis --- can be recast
to avoid elimination altogether. We will also investigate whether transfer functions
can be found, not only for sequences
of instructions, but also for entire loops~\cite{Kap05,Mon09}. Existing
approaches for the specification of (least
inductive) loop invariants rely on existential quantification~\cite[Sect.~3.4]{Mon09}, and the natural question is thus
 whether a variation of the techniques proposed in this paper can annul this complexity.

An interesting open question is whether the techniques discussed in this paper can be further generalised to
linear template constraints with variable coefficients. As discussed in Sect.~\ref{section:guards-generalisation},
the dichotomic search can be applied to any template constraint of the form $\sum_{i=1}^n c_i \cdot v_i \leq d$, where
$c_1, \ldots, c_n, d \in \mathbb{Z}$ are constants and $v_1, \ldots, v_n$ are variables. However, some
interesting abstract domains used in program analysis --- such as two variables per inequality~\cite{SKH02,simon10tvpi} ---
do not fall into this class. It is still unclear if and how such relations can be derived using binary search.
It is also interesting to note that octagons derived using our approach are tightly closed~\cite[Def.~3]{Min06}.
Intuitively, this means that all hyperplanes defined through inequalities actually touch the enclosed volume. However, the octagons may contain
redundant inequalities, which may negatively affect performance~\cite[Sect.~3.2]{BHZ09}. It will therefore be interesting
to evaluate if simplification is worthwhile~\cite[Sect.~6.1]{BHZ09} and, if so, whether
non-redundant octagons can be directly derived using SAT.

\paragraph{Acknowledgements}
This collaboration was supported by a Royal Society International Joint Project grant, reference GP101405, and by a Royal Society travel grant, reference TG092357.
The first author was supported, in part, by the DFG research training group 1298 Algorithmic Synthesis of Reactive and  Discrete-Continuous Systems and the by the DFG Cluster of Excellence on Ultra-high Speed Information and Communication, German Research Foundation grant DFG EXC 89.
The second author was funded, in part, by a Royal Society Industrial Fellowship, reference IF081178.
We gratefully acknowledge the comments provided by the reviewers of the ESOP paper~\cite{BK11a} on which this work is largely based, as well as the feedback provided by the reviewers of the SAS paper~\cite{BK10}, and its predecessor the VMCAI paper~\cite{KS10}, since it was the reviewers' critique that inspired this work.
Finally, we thank Sebastian Biallas, Stefan Kowalewski, David Monniaux, Axel Simon and Harald S{\o}ndergaard for stimulating discussions.

\bibliographystyle{abbrv}


\end{document}